\documentclass{jfm}
\usepackage{graphicx}
\usepackage{epstopdf, epsfig}
\usepackage{graphicx}
\usepackage{natbib}
\usepackage{color}
\usepackage{amsmath}
\usepackage{epstopdf}
\usepackage{color,xcolor-patch}

\ifCUPmtlplainloaded \else
\checkfont{eurm10}
\iffontfound
\IfFileExists{upmath.sty}
{\typeout{^^JFound AMS Euler Roman fonts on the system,
		using the 'upmath' package.^^J}%
	\usepackage{upmath}}
{\typeout{^^JFound AMS Euler Roman fonts on the system, but you
		dont seem to have the}%
	\typeout{'upmath' package installed. JFM.cls can take advantage
		of these fonts,^^Jif you use 'upmath' package.^^J}%
	\providecommand\upi{\pi}%
}
\else
\providecommand\upi{\pi}%
\fi
\fi

\ifCUPmtlplainloaded \else
\checkfont{msam10}
\iffontfound
\IfFileExists{amssymb.sty}
{\typeout{^^JFound AMS Symbol fonts on the system, using the
		'amssymb' package.^^J}%
	\usepackage{amssymb}%
	\let\le=\leqslant  \let\leq=\leqslant
	  \let\geq=\geqslant
}{}
\fi
\fi

\ifCUPmtlplainloaded \else
\IfFileExists{amsbsy.sty}
{\typeout{^^JFound the 'amsbsy' package on the system, using it.^^J}%
	\usepackage{amsbsy}}
{\providecommand\boldsymbol[1]{\mbox{\boldmath $##1$}}}
\fi

\providecommand\bcdot{\boldsymbol{\cdot}}

\sbox{\astrutbox}{\rule[-5pt]{0pt}{20pt}}

\newcommand{\rsx}[1]{\left.{#1}\vphantom{\Big|}\right|}

\shorttitle{Modelling double emulsion formation in planar flow-focusing microchannels}
\shortauthor{N. Wang, C. Semprebon, H. Liu, C. Zhang and H. Kusumaatmaja}

\title{Modelling double emulsion formation in planar flow-focusing microchannels}

\author{Ningning Wang \aff{1,2},
	Ciro Semprebon\aff{3}, Haihu Liu\aff{1},Chuhua Zhang\aff{1} 
	\and Halim Kusumaatmaja\aff{2}
	\corresp{\email{halim.kusumaatmaja@durham.ac.uk}}}

\affiliation{\aff{1}School of Energy and Power Engineering, Xi'an Jiaotong University, Xi'an 710049, China
	\aff{2}Department of Physics, Durham University, Durham DH1 3LE, United Kingdom
	\aff{3}3Department of Mathematics, Physics and Electrical Engineering, Northumbria University, Newcastle NE1 8ST, United Kingdom}

\begin{document}
	
	\maketitle
	
	\begin{abstract}
		Double emulsion formation in a hierarchical flow-focusing channel is systematically investigated using a free energy ternary lattice Boltzmann model. A three dimensional formation regime diagram is constructed based on the capillary numbers of the inner ($Ca_i$), middle ($Ca_m$) and outer ($Ca_o$) phase fluids. The results show that the formation diagram can be classified into periodic two-step region, periodic one-step region, and non-periodic region. By varying $Ca_i$ and $Ca_m$ in the two-step formation region, different morphologies are obtained, including the regular double emulsions, decussate regimes with one or two alternate empty droplets, and structures with multiple inner droplets contained in the continuous middle phase thread. Bidisperse behaviors are also frequently encountered in the two-step formation region. In the periodic one-step formation region, scaling laws are proposed for the double emulsion size and for the size ratio between the inner droplet and the overall double emulsion. Furthermore, we show that the interfacial tension ratio can greatly change the morphologies of the obtained emulsion droplets, and the channel geometry plays an important role in changing the formation regimes and the double emulsion sizes. In particular, narrowing the side inlets or the distance between the two side inlets promotes the conversion from the two-step formation regime to the one-step formation regime.  
	\end{abstract}
		
\section{Introduction}
		
Double emulsions are droplets with one other droplet inside. Their core-shell structure has attracted wide attentions in various fields \citep{Vladisavljevic2017}. In pharmaceuticals, one common technique is to use double emulsion for drug encapsulation of highly hydrophilic molecules. It solves the low encapsulation efficiency problem faced in single emulsion technique due to the quick partitioning of the hydrophilic molecules into the external aqueous phase \citep{Iqbal2015}. Double emulsions are also suitable containers for performing chemical and biochemical reactions under well-defined conditions. Compared to bulk reactions, the greatly reduced volume needed in double emulsion technique is beneficial for high throughput screening assays \citep{Chang2018Microfluidic}. Furthermore, double emulsions can be used as templates for the synthesis of more complex colloidosomes \citep{Azarmanesh2016,Lee2008}, as well as for controlled release of the inner contents \citep{Datta2014}. To ensure the successful applications of double emulsions, one of the key issues is to provide precise control over the double emulsion structure, size and monodispersity at a sufficient production rate \citep{Shang2017, Zhang2016Controllable}.

Traditional double emulsion fabrication methods, such as the bulk emulsification and the membrane emulsification methods \citep{Vladisavljevic2017}, are attractive to many industries (e.g. food and cosmetic) where scalability for large production is important \citep{Varka2012}. However, these techniques have poor size and monodispersity control \citep{Silva2016}, which makes them inadequate for applications requiring high precision, such as in medical, pharmaceutical, and material industries. The emergence of microfluidic technology \citep{Utada2005, Whitesides2006} opens up a new horizon. It provides more detailed control over the operating conditions \citep{Vladisavljevic2017} and offers great flexibility in designing multi-layer \citep{Abate2009} or multi-core emulsions \citep{Nisisako2005, Nabavi2017M}. So far, the microfluidic devices for producing double emulsions can be roughly classified into a series of two T-junctions \citep{Okushima2004}, two flow-focusing junctions \citep{Pannacci2008,Abate2011,Seo2007}, co-axial capillaries \citep{Utada2005,Nabavi2017M}, and the possible combinations and variations of the aforementioned geometries \citep{Nisisako2016,Zhu2016}. 

The understanding of double emulsion formation dynamics are crucial for microfluidic control and equipment optimization. Double emulsions are commonly generated either in a two-step or one-step formation regime, depending on whether the inner part of the double emulsion is sheared simultaneously with the middle layer in the outer fluid \citep {Utada2005}. Due to the distinct flow details in the two-step and one-step formation regimes, the influence of flow conditions, fluid properties and geometrical parameters on each regime should be analyzed respectively. For the two-step formation regime, \citet{Okushima2004} have systematically showed the effect of flow rates on the double emulsion sizes and the number of internal droplet for multi-core emulsions when they are formed using a series of T-junctions. The one-step formation regime is mostly encountered in co-axial microcapillary devices. Experimental studies have been carried out on the effect of flow rates \citep{Lee2008,Kim2013} and geometrical settings \citep{Nabavi2017}. Scaling laws have also been developed for the emulsion size predictions \citep{Utada2005, Chang2009}. 

Complementary to experiments, numerical studies on double emulsion formation dynamics in microfluidic channels have also garnered strong interest. For instance, great efforts were made to elucidate the effects of flow rates, interfacial tension ratios, geometry \citep{Chen2015, Nabavi2015}, and viscosities \citep{Fu2016} on the double emulsion properties and the flow regime predictions \citep{Nabavi2017} for co-axial flow-focusing capillary devices. Simulations are particularly advantageous for providing accurate flow details and for allowing each relevant parameter in the system to be varied systematically. In the literature, a number of ternary fluid models have been successfully developed and applied in the study of multiple emulsion formation behaviors, including using the volume of fluid (VOF) method \citep{Chen2015, Nabavi2015, Nabavi2017, Azarmanesh2016}, the front-tracking method \citep{Vu2013}, the free energy finite element method \citep{Park2012} and the lattice Boltzmann method \citep{Fu2016, Fu2017}. 

In this work, our focus is on the planar flow-focusing cross-junctions. They are promising for integration with other devices and they can be parallelized to raise the production rate of the emulsion droplets, while still ensuring good size control \citep{Lee2016}. Furthermore, in contrast to other microfluidic geometries, systematic parametric study is rarely reported on planar flow-focusing devices. Several works, such as \citet{Abate2011} and \citet{Azarmanesh2016}, briefly discussed the possible conversion between the two-step and one-step formation regimes and the variation of shell thickness. However, it remains unclear in which flow rate regions monodisperse double emulsions are produced; and correspondingly, how  the droplet sizes can be varied in those regions. It is likely that the droplet sizes have different dependencies on the flow rates for the two-step and one-step formation processes. There are also open questions on the role of channel geometry in the formation regime conversion, and on the effects of interfacial tension ratio in determining the morphologies of the emulsion droplets, including the possibility of complete, partial and non-engulfing shapes \citep{Guzowski2012, Pannacci2008, Chao2016, Nisisako2010}. 

We have chosen to use the lattice Boltzmann method (LBM). LBM is highly favorable for the study of emulsion formation behaviors due to its simplicity in solving interface dynamics, including droplet break-ups and coalescences, as well as its ability to deal with complex geometries, and its high efficiency in parallel computation \citep{Kruger2017}. So far, three types of ternary lattice Boltzmann models have been developed, including the free energy model \citep{Semprebon2016, Liang2016, Wohrwag2018, Abadi2018}, color-fluid model \citep{Leclaire2013a, Leclaire2013b, Yu2019,Fu2016,Fu2017}, and the Shan-Chen type models \citep{Bao2013, Wei2018}. 

Here we improve on the free energy lattice Boltzmann model developed by \citet{Semprebon2016}. A major progress is that our model allows a wider range of interfacial tension ratios, such that all possible biphasic emulsion morphologies can be captured \citep{Guzowski2012}, including complete and non-engulfing shapes. The model by \citet{Semprebon2016} only allows partial engulfing shapes. Coupling the free energy model with the advantages of the lattice Boltzmann method, we conduct a systematic study on the dynamics of double emulsion formation behaviors in planar hierarchical flow-focusing junctions. We focus on the two-dimensional (2D) case to reduce the computational time needed for parametric studies. The major physical difference in the flow dynamics between the 2D and the three-dimensional (3D) systems lies in the lack of an additional Laplace pressure induced by the out-of-plane curvature \citep{Chen2017}. Such contribution can accelerate the droplet pinch-off process \citep{Hoang2013}, especially at large Weber number. However, we believe  most of the fundamental flow physics are still involved in the 2D system and a systematic 2D study can still capture some of the key qualitative trends in the formation regimes and emulsion sizes as function of the flow rates of each fluid phase.

The paper is organized as follows. In \S 2, we describe the improved ternary free energy model, the lattice Boltzmann method, and the boundary conditions involved. In \S 3, we validate the model and boundary conditions by Poiseuille flow, moving droplet and static emulsion morphology tests. In \S 4, our systematic parametric study allows us to construct a flow regime diagram, where we describe a wide range of formation regimes, including the periodic two-step and one-step double emulsion formation regimes, decussate regime, bidisperse regime and even the continuous structure with multiple inner droplets. Scaling laws are also proposed for the double emulsions produced in the one-step formation regime, and the effects of the interfacial tension ratios and the geometrical parameters are investigated. Finally, we summarize our main findings and forecast prospects for future work in \S 5.
	
\section{Numerical method}
\subsection{Free-energy model}
		
The present model is developed based on the equal-density ternary free-energy lattice Boltzmann model proposed by \citet{Semprebon2016}. 
The system is described by the free-energy functional 
	\begin{equation}
	\mathcal{H}=  \int_\Omega c_s^2\rho \ln \rho \,\, dV + \mathcal{F}.
	\end{equation}
The first term is the standard ideal gas term in the lattice Boltzmann method with $c_s={1/\sqrt{3}}$ and $\rho$ the total density. $\Omega$ is the system volume. This term does not affect the phase behaviour. To realise three fluid components, the second term $\mathcal{F}$ is introduced and it is given by
	\begin{eqnarray}
	\mathcal{F}&=&\sum_{m=1}^3\int_\Omega (E_0(C_m)+E_{interface}(\nabla C_m)) dV, \nonumber\\
	&=& \sum_{m=1}^3\int_\Omega [\frac{\kappa_m}{2}C_m^2(1-C_m)^2
	+\frac{\alpha^2\kappa_m}{2}(\nabla C_m)^2] dV,
	\label{eq:free-energy}
	\end{eqnarray}
It is constructed using a double-well potential form for the bulk free energy $E_0(C_m) = (\kappa_m/2)C_m^2(1-C_m)^2$ and a square gradient term for the interface region $E_{interface}(\nabla C_m) = (\alpha^2\kappa_m/2)(\nabla C_m)^2$. $C_m$ ($m$ = 1,2,3) are the concentration fractions with two minimum values at $C_m=0$ and $C_m=1$ for each component $m$. In the current model, all components have the same density $\rho_m$, which we have set to be 1.0 for simplicity. Thus the total density is related to the concentration fractions by defining $\rho = C_1\rho_1 + C_2\rho_2 + C_3\rho_3 = C_1 + C_2 + C_3$, which is equal to 1.0 in this model. Three physically meaningful bulk phases termed red, green and blue here could be denoted by $\mathsfbi{C}$ = $[C_1 \quad C_2 \quad C_3]$ = $[1 \quad 0 \quad 0]$, $[0 \quad 1 \quad 0]$ and $[0 \quad 0 \quad 1]$, respectively. $\alpha$ is the interface width parameter. The interfacial tension between red-green phases $\sigma_{rg}$, red-blue phases $\sigma_{rb}$, and green-blue phases $\sigma_{gb}$ are related to $\kappa$'s through 
	\begin{eqnarray}
	\sigma_{rg} &=&\frac{\alpha}{6}(\kappa_1+\kappa_2),      \nonumber\\
	\sigma_{rb} &=&\frac{\alpha}{6}(\kappa_1+\kappa_3),      \nonumber\\
	\sigma_{gb} &=&\frac{\alpha}{6}(\kappa_2+\kappa_3).   
	\label{eq:sigma}
	\end{eqnarray}

To capture the interface dynamics, two order parameters $\phi$ and $\psi$ are introduced as 
	\begin{eqnarray}
	\quad \phi = C_1 - C_2, \quad \psi = C_3, 
	\label{eq:orderparameter}
	\end{eqnarray}
and the original concentration fields can be reversely obtained from $\rho$, $\phi$ and $\psi$ via $C_1 = (\rho + \phi - \psi)/2$, $C_2 = (\rho - \phi - \psi)/2$, and $C_3 = \psi$. The order parameters and the hydrodynamics of the ternary fluid system are governed by two Cahn-Hilliard equations, the continuity and the Navier-Stokes equations: 
	\begin{eqnarray}
	\partial_t \phi  +  \nabla\bcdot(\phi \boldsymbol{v}) &=& M_\phi \nabla^2 \mu_{\phi},
	\label{eq:ch-phi} \\
	\partial_t \psi  +  \nabla\bcdot(\psi \boldsymbol{v}) &=& M_\psi \nabla^2 \mu_{\psi},
	\label{eq:ch-psi} \\
	\partial_t \rho +  \nabla\bcdot(\rho \boldsymbol{v})  &=& 0, \label{eq:mass}\\
	\partial_t (\rho \boldsymbol{v}) + \nabla\bcdot( \rho \boldsymbol{v} \boldsymbol{v}) & = &  \nabla\bcdot[\eta (\nabla \boldsymbol{v} + \nabla \boldsymbol{v}^T)] -  \nabla\bcdot\mathsfbi{P}.
	\label{eq:momentum}
	\end{eqnarray}
Here, $ \boldsymbol{v} $ is the fluid velocity and $\eta$ is the dynamic viscosity. $ M_\phi $ and $ M_\psi $ are the mobility values for $\phi$ and $\psi$. The thermodynamic properties are related to the equations of motion via the chemical potential $\mu_\phi$, $\mu_\psi$ and the pressure tensor $\mathsfbi{P}$. The chemical potential is defined as the variational derivative of the free energy $\mu_q = \delta \mathcal{F}/\delta q$, where $q = \rho, \phi$ or $ \psi $. The pressure tensor term in Eq. (\ref{eq:momentum}) is constructed by $\nabla\bcdot\mathsfbi{P} = \nabla (\rho c_s^2) + \nabla\bcdot\mathsfbi{P_0}$ and $\nabla\bcdot\mathsfbi{P_0} = (\rho \nabla \mu_\rho + \phi \nabla \mu_\phi  + \psi \nabla \mu_\psi) $. The first term $\nabla (\rho c_s^2)$ is the standard ideal gas term in LBM and it is simply incorporated in the equilibrium distribution function \citep{Briant2004,Zhang2011lattice}. The $\boldsymbol{F}= - \nabla\bcdot\mathsfbi{P_0}$ term is treated as an external force term in the lattice Boltzmann implementation. The explicit expressions of $\mu_\rho$, $\mu_\phi$, $\mu_\psi$, and $\mathsfbi{P}$ are given in \citet{Semprebon2016}.  
\begin{figure}
	\centerline{\includegraphics[scale=0.5]{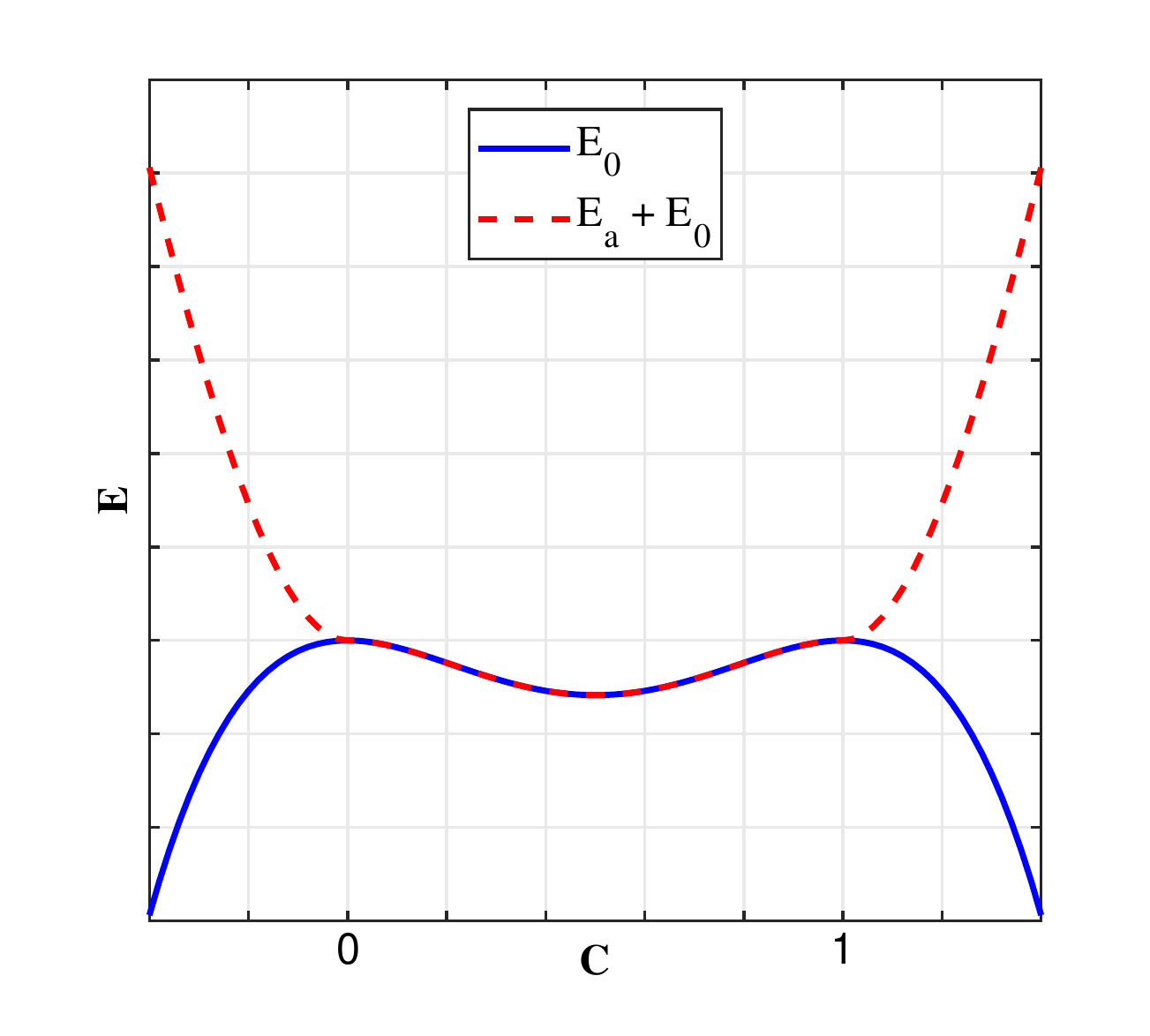}}
	\caption{Illustration of the bulk free energy profile without (blue solid line) and with (red dashed line) the additional free energy term $E_a$, given by Eq. (\ref{eq:Ea}), for a negative $\kappa_m$.} 
	\label{fig:free_energy}
\end{figure}

Consider now a case where a red droplet is completely engulfed by a green one and they are submerged in a blue phase fluid at thermodynamic equilibrium. According to the theoretical analysis of \citet{Guzowski2012}, the interfacial tensions should satisfy $\sigma_{rb}>\sigma_{gb}+\sigma_{rg}$. Given the relation between the $\sigma$'s and $\kappa_m$'s in Eq.(\ref{eq:sigma}), one can easily find that $\kappa_2$ should be negative while $\kappa_1$ and $\kappa_3$ are positive. In the free energy model, the negative $\kappa_2$ will invert the shape of the bulk free energy profile $E_0(C_m)$: the two minimum values at 0 and 1.0 become two maximum values as shown by the blue solid line in figure \ref{fig:free_energy}. As such, setting one of the $\kappa_m$'s to be negative often leads to incorrect results or even numerical instability as the concentration value deviates significantly from [0, 1.0]. A similar situation has been encountered in other LB models, and a simple remedy has been proposed by introducing an additional free energy term, see \citet{Lee2010} and \citet{Abadi2018}. Inspired by these works, to solve the problem induced by negative $\kappa_m$, here we introduce an additional free energy term given by
	\begin{equation}
	E_a(C_m) = \left\{
	\begin{array}{ll}
	\beta C_m^2, & C_m < 0 \\[2pt]
	0,         & 0 \le C_m \le 1 \\[2pt]
	\beta (C_m - 1)^2,  & C_m > 1,        
	\end{array} \right.
	\label{eq:Ea}
	\end{equation}
where $\beta$ is an adjustable positive parameter controlling the slope of the energy profile $E_0 + E_a$ as depicted by the red dashed line in figure \ref{fig:free_energy}. Since we add a new free energy term in Eq. (\ref{eq:Ea}), additional terms should be included in the chemical potentials accordingly, which are listed in Appendix \ref{appA}.	
		
\subsection{Lattice Boltzmann method}
		
To solve Eq. (\ref{eq:ch-phi})-(\ref{eq:momentum}) using the lattice Boltzmann method, three distribution functions are introduced: $f_i({\boldsymbol r}, t)$ for the density, and $g_i({\boldsymbol r}, t)$ and $h_i({\boldsymbol r}, t)$ for the order parameters $\phi$ and $\psi$, respectively. The distribution functions are discretized in space ${\boldsymbol r}$ and time $t$, according to a set of lattice velocity vectors $\boldsymbol{e}_i$. In the D2Q9 discrete scheme (two-dimension with nine discrete velocities) used here, the lattice velocities are given as $\boldsymbol{e}_0$ = $(0,0)$, $\boldsymbol{e}_{1,3} = (\pm 1,0)$, $\boldsymbol{e}_{2,4} = (0,\pm 1)$, $\boldsymbol{e}_{5,7} = (\pm 1,\pm 1)$ and $\boldsymbol{e}_{6,8} = (\mp 1,\pm 1)$, as shown in figure \ref{fig:moving_droplet}(a). The time evolution of the distribution functions includes the collision and streaming steps, which can be written as
	\begin{eqnarray}
	f_i({\boldsymbol r + \boldsymbol{e}_i \delta_t}, t + \delta_t) &=& f_i({\boldsymbol r}, t) - \frac{1}{\tau_f} [f_i({\boldsymbol r}, t)-f_i^{eq}(\rho, \widetilde{\boldsymbol{u}})] \nonumber\\
	&& + [f_i^{eq}(\rho, \widetilde{\boldsymbol{u}}+\delta \widetilde{\boldsymbol{u}}) - f_i^{eq}(\rho, \widetilde{\boldsymbol{u}}) ], 
	\label{eq:f-function} \\
	g_i({\boldsymbol r + \boldsymbol{e}_i \delta_t}, t + \delta_t) &=& g_i({\boldsymbol r}, t) - \frac{1}{\tau_g} \left[g_i({\boldsymbol r}, t)-g_i^{eq}({\boldsymbol r}, t)\right], 
	\label{eq:g-function} \\
	h_i({\boldsymbol r + \boldsymbol{e}_i \delta_t}, t + \delta_t) &=& h_i({\boldsymbol r}, t) - \frac{1}{\tau_h} \left[h_i({\boldsymbol r}, t)-h_i^{eq}({\boldsymbol r}, t)\right]. 
	\label{eq:h-function}
	\end{eqnarray}
Here, the force term is implemented through the exact difference method \citep{Kuper2009,Mazloomi2015}, which is expressed as the last two terms enclosed in brackets in Eq. ($\ref{eq:f-function}$), with $\widetilde{\boldsymbol{u}} = \sum_i f_i \boldsymbol{e}_i/\rho $, i.e., the velocity without the force term, and $\delta \widetilde{\boldsymbol{u}} = \boldsymbol{F} \delta t/\rho$. The lattice time step $\delta_t$ is set to be 1.0. $\tau_f$ is the relaxation parameter given by $1/\tau_f = C_1/\tau_{1} + C_2/\tau_{2} + C_3/\tau_{3}$, where $\tau_{1, 2, 3}$ are related to the viscosity of each fluid by $\tau_{1, 2, 3} = 3\eta_{r,g,b}/\rho + 1/2$, respectively \citep{Kruger2017}. $\tau_g$ and $\tau_h$ are the relaxation parameters that are related to the mobility parameters $M_\phi$ and $M_\psi$ in the Cahn-Hilliard equations through
	\begin{eqnarray}
	&M_\phi \ =\ \Gamma_\phi (\tau_g - \frac{\delta t}{2}),\ \\ 
	&M_\psi \ =\ \Gamma_\psi (\tau_h - \frac{\delta t}{2}),\
	\end{eqnarray}
where $\Gamma_\phi$ and $\Gamma_\psi$ are two tunable parameters. The mobility values $M_\phi$ and $M_\psi$ are relevant for the timescale of Cahn-Hilliard diffusion and the relaxation time of the interface. Generally, the mobility values should be sufficiently large to retain the interfacial thickness, but small enough to insure the reasonable damping of the convective term \citep{Lim2014,Jacqmin1999}. At present it remains an open problem to assign mobility values in numerical studies. Indeed most papers use comparison with experiments to set the values, and one common solution is to use mobility related dimensionless parameters, e.g., the Peclet number ($Pe$). In our microfluidic study, a characteristic $Pe$ is defined based on the middle phase fluid as $Pe_m=(u_m w_1)/(M_{\phi}\kappa_2)$. The absolute values of $Pe_m$ used is generally on the order of $o(10)-o(80)$, which is on similar magnitudes to those used in previous two-phase droplet behavior studies \citep{Shardt2014,Zhou2010,Menech2006}. Moreover, $M_\psi=M_\phi/3$ is considered to assign symmetric mobility for each concentration component \citep{Semprebon2016}.

$f_i^{eq}$, $g_i^{eq}$ and $h_i^{eq}$ are the local equilibrium distribution functions, which are given by
	\begin{eqnarray}
	f_i^{eq} &=& \omega_i \rho [1 + 3 \boldsymbol{e}_i \cdot \widetilde{\boldsymbol{u}} + \frac{9}{2}(\boldsymbol{e}_i \cdot \widetilde{\boldsymbol{u}})^2 -  \frac{3}{2} \widetilde{\boldsymbol{u}}^2], 
	\label{eq:feq} \\	
	g_i^{eq} &=& \left\{
	\begin{array}{ll}
	\omega_i [3 \Gamma_\phi \mu_\phi + 3 \phi \boldsymbol{e}_i \cdot \boldsymbol{v} + \frac{9\phi}{2} (\boldsymbol{e}_i \cdot \boldsymbol{v})^2 -  \frac{3\phi}{2} \boldsymbol{v}^2], & i = 1-8 \\[2pt]
	\phi - \sum\limits_{i=1}^{8}g_i^{eq},         & i = 0        
	\end{array} \right.
	\label{eq:geq} \\
	h_i^{eq} &=& \left\{
	\begin{array}{ll}
	\omega_i [3 \Gamma_\psi \mu_\psi + 3 \psi \boldsymbol{e}_i \cdot \boldsymbol{v} + \frac{9\psi}{2} (\boldsymbol{e}_i \cdot \boldsymbol{v})^2 -  \frac{3\psi}{2} \boldsymbol{v}^2], & i = 1-8 \\[2pt]
	\psi - \sum\limits_{i=1}^{8}h_i^{eq},         & i = 0        
	\end{array} \right.
	\label{eq:heq}
	\end{eqnarray}
where the weight coefficient $\omega_i$ are given by $\omega_0=4/9$, $\omega_{1-4}=1/9$ and $\omega_{5-8}=1/36$. The macroscopic variables are related to the distribution functions through
	\begin{equation}
	\rho=\sum_i f_i , \;\;\;\;\;\; 
	\rho \boldsymbol{v} = \sum_i f_i \boldsymbol{e}_i + \frac{\boldsymbol{F} \delta t}{2} ,  \;\;\;\;\;\; 
	\phi = \sum_i g_i, \;\;\;\;\;\; 
	\psi = \sum_i h_i. \;\;\;\;\;\; 
	\label{eq:LBmoments}
	\end{equation}

\subsection{Boundary conditions}
	
The boundary conditions involved in the present study contain: no-slip boundary, wetting boundary and the inlet-outlet boundary. No-slip boundary condition is used on the solid walls, which is realized by the half-way bounceback rule \citep{Ladd1994}. The solid walls should have a preferential affinity with the continuous phase fluid to generate stable droplets/emulsions \citep{Abate2011}. \citet{Fu2016} successfully implemented the wetting boundary condition by setting a fictive density on the walls in a LB ternary color-fluid model. Similarly for the free energy model used here, the macroscopic values of $\rho$, $\phi$ and $\psi$ on the walls are designated to be the same as those of the continuous phase fluid that is assumed to completely wet the walls. For the velocity inlet, the Zou-He velocity boundary condition \citep{Zou1997} is applied to solve the unknown density distribution functions of $f_i$. To obtain the unknown $g_i$ and $h_i$ values at the inlet, the method used by \citet{Hao2009} and \citet{Liu2011} is adopted. Take figure \ref{fig:moving_droplet}(a) for instance, given an inlet boundary with the inflow direction pointing to the right, $g_{1,5,8}$ and $h_{1,5,8}$ are unknown after the streaming step. We assume that one pure single fluid exists at the inlet, where the prescribed values of $\phi$ and $\psi$ are $\phi_{in}$ and $\psi_{in}$, respectively. The sum of the unknown distribution functions can be solved according to Eq. (\ref{eq:LBmoments}), and then $g_{1,5,8}$ and $h_{1,5,8}$ are allocated by their weight factors as
\begin{eqnarray}
\rsx{g_i}_{i=1,5,8} &=& \frac{g^*\omega_i}{\omega_1 + \omega_5 + \omega_8},  \;\;\;\;\;\; 
g^* = g_1 + g_5 + g_8 = \phi_{in} -  \sum_i \rsx{g_i}_{i=0,2,3,4,6,7},  
\label{eq:inletgi} \\
\rsx{h_i}_{i=1,5,8} &=& \frac{h^*\omega_i}{\omega_1 + \omega_5 + \omega_8},  \;\;\;\;\;\; 
h^* = h_1 + h_5 + h_8 = \psi_{in} -  \sum_i \rsx{h_i}_{i=0,2,3,4,6,7}.  
\label{eq:inlethi}
\end{eqnarray}

For the outlet boundary, the convective boundary condition (CBC) \citep{Lou2013,Chen2017} is used for its good performance in multicomponent flow simulations. In the present model, the CBC is harnessed in two aspects. One is for the unknown distribution functions $\chi_i$ = $f_i$, $g_i$ and $h_i$ at the outlet layer ($x=L_x$),
\begin{equation}
	\chi_i(L_x,y,t+\delta t)  = \frac{\chi_i(L_x,y,t) + \zeta(L_x-1,y,t)\chi_i(L_x-1,y,t+\delta t)}{1+\zeta(L_x-1,y,t)}.
\label{eq:cbc1} 
\end{equation}
The other is for the macroscopic quantities, such as $\chi'$ = $\rho$, $\phi$, $\psi$ and $\mathsfbi{P_0}$ on the ghost layer right outside the outlet, i.e., $x=L_x+1$, which is needed to compute the derivative terms at the outlet fluid layer, 
\begin{equation}
	\chi'(L_x+1,y,t+\delta t)  = \frac{\chi'(L_x+1,y,t) + \zeta(L_x,y,t)\chi'(L_x,y,t+\delta t)}{1+\zeta(L_x,y,t)}.
\label{eq:cbc2}
\end{equation}
Here, $\zeta$ is the characteristic velocity normal to the outlet boundary. For simplicity, we have explicitly computed $\chi_i$ and $\chi'$ through $\zeta$ at time $t$. Three common choices for $\zeta$ in CBC's are the average velocity (CBC-AV), local velocity (CBC-LV) and the maximum velocity (CBC-MV) \citep{Lou2013}.	
	
\section{Model validation}

\subsection{Convective outlet boundary conditions}
	
In this section, the performance of the CBC in the present model is tested by simulating a single-phase Poiseuille flow and a Poiseuille flow with a moving droplet. In the single-phase Poiseuille flow settings, a fluid with viscosity of 0.167 flows in the $x$ direction with a maximum velocity of $u_{max}$ = 0.0015 in a computational domain of $L_x \times L_y = 99 \times 39$. No-slip boundaries are used both on the top and bottom walls. The Zou-He velocity inlet is applied with a parabolic velocity distribution given as
	\begin{equation}
	u_x(y) = \frac{-4u_{max}(y-y_1)(y-y_2)}{(y_2 - y_1)^2}, \quad 1\leq y\leq L_y-1,
	\label{eq:anavel}
	\end{equation}
	\begin{figure}
	\centering{\includegraphics[scale=0.5]{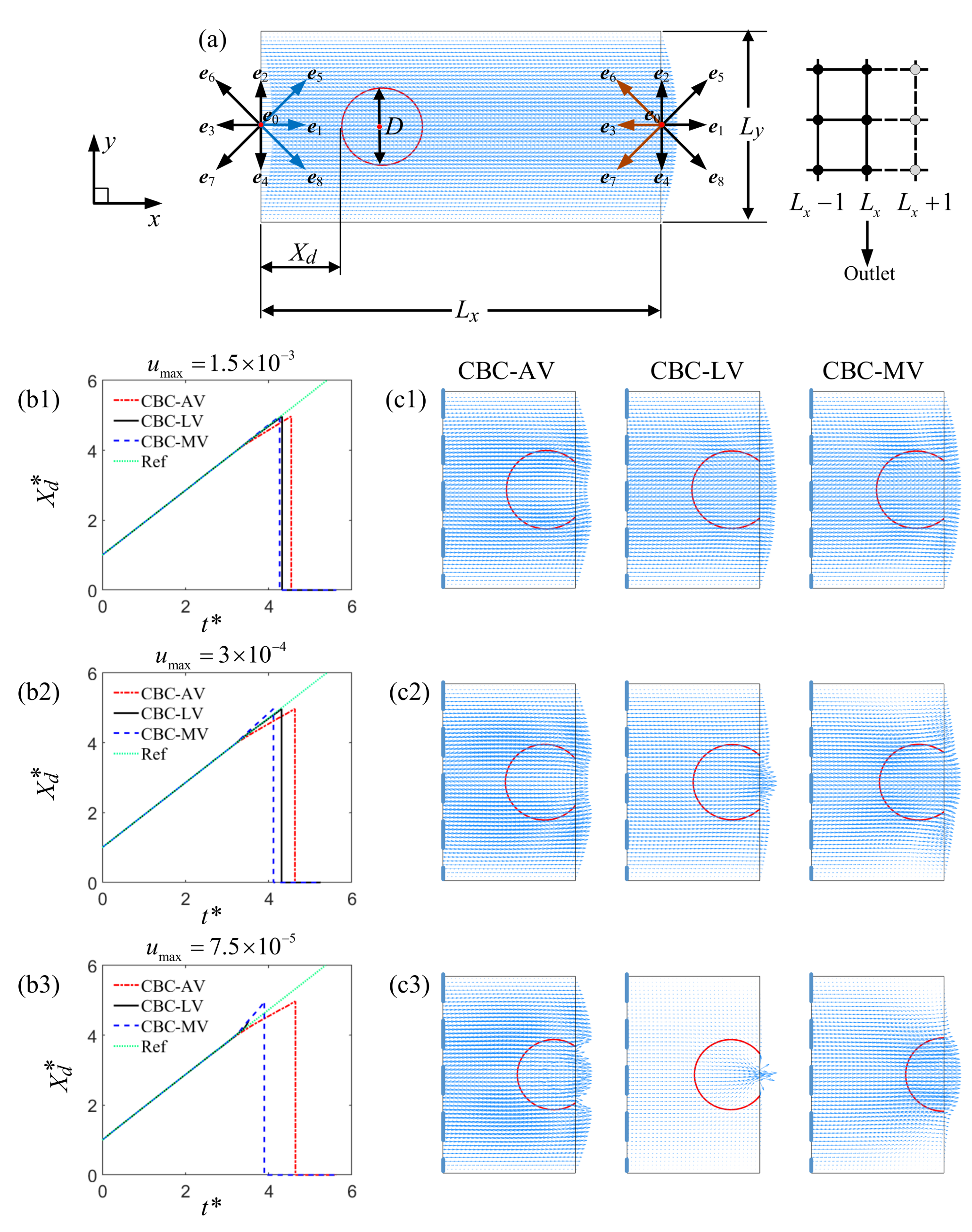}}
	\caption{(a) Illustration of the moving droplet setup; (b1-b3) time histories of $X_d$ for $u_{\rm max} = 1.5 \times 10^{-3}$, $3.0 \times 10^{-4}$ and $7.5 \times 10^{-5}$ with different CBC's; (c1-c3) typical snapshots of the droplet shape and velocity vectors when the droplet passes through the outlet layer with each CBC boundary at $u_{\rm max}$ corresponding to (b1-b3), respectively. The $X_d$ and time values are normalized using $X^*_d = X_d/D$ and $t^* = tu_{max}/D$.}
	\label{fig:moving_droplet}
	\end{figure}
where $y_1 = 0.5$ and $y_2 = L_y - 0.5$ are the locations of the bottom and top walls. All three options of the CBC mentioned above are implemented at the right outlet, and their accuracy is quantified using the relative velocity error computed by $E_u = \sqrt{\sum((u_x)_{ana} - (u_x)_{simu})^2/\sum((u_x)_{ana}^2)}$, where $(u_x)_{ana}$ is the analytical velocity given by Eq. (\ref{eq:anavel}) and $(u_x)_{simu}$ denotes the simulated velocity. The obtained values of $ E_u $ under CBC-AV, CBC-LV and CBC-MV conditions are $1.449 \times 10^{-4}$, $1.151 \times 10^{-4}$ and $2.254 \times 10^{-4}$ in the middle of the channel, i.e., $ x = 49 $, and  $1.454 \times 10^{-4}$, $5.4 \times 10^{-3}$ and $3.144 \times 10^{-4}$ at the outlet layer. It is seen that all three outlet boundaries give satisfactory results for flow far away from the outlet. However, the accuracy at the outlet layer varies: the CBC-AV provides the highest accuracy, CBC-MV is slightly lower and CBC-LV shows the poorest performance.
	
In the moving droplet test, a droplet with radius $R$ = 20 is centered at $(60, 49.5)$ in a channel of $L_x \times L_y = 199 \times 99$, as illustrated in figure \ref{fig:moving_droplet} (a). The two fluid phases have the same viscosity of 0.167 and their interfacial tension $\sigma$ is 0.005. All the boundary conditions are the same as those in the single-phase Poiseuille flow simulations. The whole fluid domain is initialized with a uniform parabolic velocity profile as given by Eq. (\ref{eq:anavel}). Three different values of $u_{max}$ are tested, i.e., $u_{max}$ = $1.5 \times 10^{-3}$, $3.0 \times 10^{-4}$ and $7.5 \times 10^{-5}$. To make a quantitative comparison, the time history of the distance $X_d$ measured from the inlet to the leftmost point of the droplet is recorded and shown in figure \ref{fig:moving_droplet} (b1-b3). The $X_d$ and time are normalized using $X^*_d = X_d/D$ and $t^* = tu_{max}/D$, where $D$ is the droplet diameter. The $X^*_d$ curve of the droplet moving in a longer channel ($L_x \times L_y = 399 \times 99$) computed with CBC-AV is used as the reference result for each flow condition. Note in figure \ref{fig:moving_droplet} (b1-b3) that the sharp decrease of $X^*_d$ occurs when the droplet completely moves out of the channel. 

It is seen in figure \ref{fig:moving_droplet} (b1-b3) that the $X^*_d$ increases linearly with time and agrees with the reference line before the droplet interface touches the outlet boundary for each of the tested flow conditions. The option of the CBC's has little effect on the flow behaviors away from the outlet. Deviations in $X^*_d$ curves from the reference lines occur at around $t^*$ = 4 when the droplet passes through the outlet. Compared to the reference lines, the case with CBC-AV slightly lags behind, and the case with CBC-MV moves a bit faster. Also, the case with CBC-LV gives the most accurate results for moderate characteristic velocities, as illustrated in figure \ref{fig:moving_droplet} (b1)-(b2). The deviation in $X^*_d$ increases as $u_{max}$ decreases for the cases with CBC-AV and CBC-MV. When the $u_{max}$ is on the same order of magnitude as the spurious velocities of the present model, i.e., $u_{max}=7.5 \times 10^{-5}$ in (b3), numerical instability arises for the case with CBC-LV, whereas the cases with CBC-AV and CBC-MV show better robustness. Due to the low velocity often encountered in double emulsion generation, the robustness of the outlet boundary at low velocities is of great significance. On the other hand, for low velocity cases shown in (c2)-(c3), the velocity in regions close to the walls is less affected for the case with CBC-AV than that with CBC-MV. The momentum deficit or surplus around the outlet region could be attributed to the momentum imbalance at the outlet, which is not fully ensured by the CBC when the external force term is solved in the potential form \citep{Li2017}. The resulting velocity profile is also affected by the form of the characteristic velocity used in the CBC. Considering all the above tests, CBC-AV generally shows better performance and it is therefore used in the following studies.

In addition, since we find the flow behaviours are unaffected away from the outlet, we always use channel length which is much larger compared to the typical emulsion droplet, in order to minimise any undesirable effect from the outlet boundary condition.
\begin{figure}
	\centering{\includegraphics[scale=0.5]{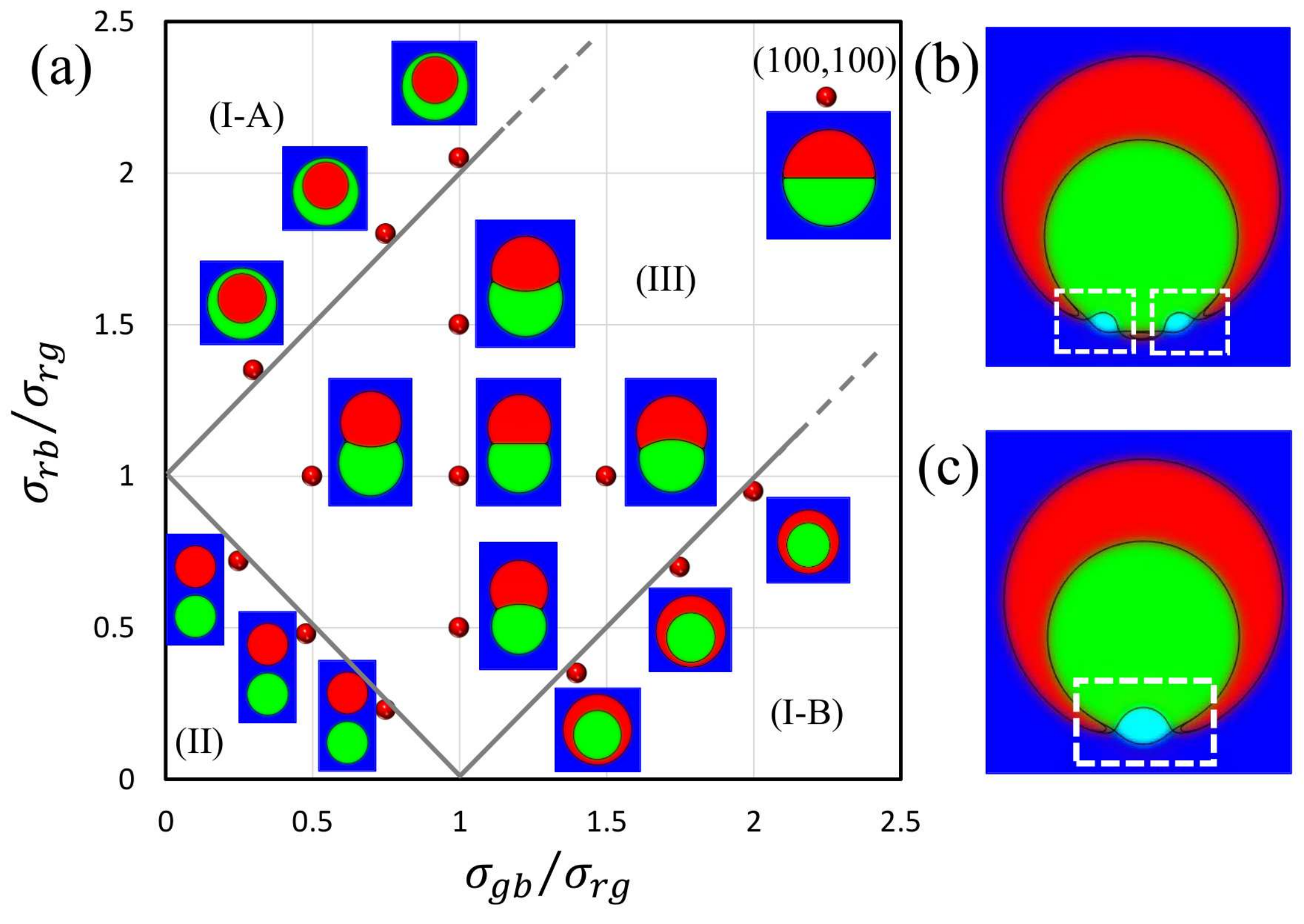}}
	\caption{(a) Morphology diagram for two equal sized droplets with $\beta = 0.001$; (b) emulsion shape with $\beta = 0$ for ($\sigma_{gb}/\sigma_{rg}$, $\sigma_{rb}/\sigma_{rg}$) = (1.4, 0.35); (c) emulsion shape with $\beta = 0.0001$ for ($\sigma_{gb}/\sigma_{rg}$, $\sigma_{rb}/\sigma_{rg}$) = (1.4, 0.35).}
	\label{fig:stability_diagram}
\end{figure}

\subsection{Morphology diagram}

Since the interfacial tension relations are crucial in determining ternary emulsion morphologies \citep{Pannacci2008,Guzowski2012}, another validation test is conducted to show the capability of the current model in simulating a wide range of interfacial tension ratios. Following the theoretical analysis of \citet{Guzowski2012}, two equal-sized red and green droplets are initially put next to each other and dispersed in the outer blue fluid. Three typical thermodynamic equilibrium morphologies can be obtained depending on the interfacial tension ratios of $\sigma_{gb}/\sigma_{rg}$ and $\sigma_{rb}/\sigma_{rg}$, as divided by the solid lines shown in figure \ref{fig:stability_diagram} (a): (I-A) complete engulfing with the red droplet inside the green one for $\sigma_{rb}/\sigma_{rg} > 1 + \sigma_{gb}/\sigma_{rg}$; (I-B) complete engulfing with the green droplet inside the red one for $\sigma_{gb}/\sigma_{rg} > 1 + \sigma_{rb}/\sigma_{rg}$; (II) non-engulfing, for $\sigma_{rb}/\sigma_{rg} + \sigma_{gb}/\sigma_{rg} < 1$, where the red and green droplets tend to separate from each other; (III) partial engulfing (Janus droplet), for $|(\sigma_{rb}/\sigma_{rg}) - (\sigma_{gb}/\sigma_{rg})|<1$ and $\sigma_{rb}/\sigma_{rg} + \sigma_{gb}/\sigma_{rg} > 1$, where the interfacial tensions satisfy a Neumann triangle. 

In our numerical test, the red and green droplets are both initialized with radius $R = 60$ surrounded by the blue fluid in a domain of $L_x \times L_y = 399 \times 399$. All the fluid viscosities are 0.167. The initial concentration fractions for three fluids are given by \citep{Yu2019}
\begin{eqnarray}
C_1(x,y) &=& 0.5 + 0.5 \tanh \left[ \frac{R-\sqrt{(x-199.5)^2+(y-199.5-R)^2}}{2\alpha} \right],  \\
\label{eq:sb_c1}
C_2(x,y) &=& 0.5 + 0.5 \tanh \left[ \frac{R-\sqrt{(x-199.5)^2+(y-199.5+R)^2}}{2\alpha} \right],  \\
\label{eq:sb_c2}
C_3(x,y) &=& 1 - C_1(x,y) - C_2(x,y).
\label{eq:sb_c3}
\end{eqnarray}  

Periodic boundary conditions are used for all boundaries. To reproduce all the possible morphologies, simulations are performed at various groups of ($\sigma_{gb}/\sigma_{rg}$, $\sigma_{rb}/\sigma_{rg}$): (I-A) complete engulfing with red droplet inside: (0.3, 1.35), (0.75, 1.8), (1.0, 2.05); (I-B) complete engulfing with green droplet inside: (1.4, 0.35), (1.75, 0.7), (2.0, 0.95); (II) non-engulfing: (0.48, 0.48), (0.25, 0.72), (0.75, 0.23); and (III) partial engulfing emulsions: (1.0, 1.0), (1.0, 1.5), (1.0, 0.5), (0.5, 1.0), (1.5, 1.0) and (100, 100). The interfacial tension $\sigma_{rg}$ is fixed at 0.005 except for the case with ($\sigma_{gb}/\sigma_{rg}$, $\sigma_{rb}/\sigma_{rg}$) = (100, 100), where $\sigma_{rg}$ = 0.00001 is used to reach the high interfacial tension ratio. The value of the coefficient $\beta$ is set to be 0.001 for the additional free energy term. The simulated equilibrium morphologies are shown by the insets in figure \ref{fig:stability_diagram} (a). Good agreements with theoretical morphologies are achieved for all types of emulsions. Moreover, \citet{Pannacci2008} experimentally investigated the equilibrium states of compound emulsions. Their results are presented as a function of the spreading coefficients, i.e., $S_i = \sigma_{jk}-\sigma_{ij}-\sigma_{ik}$ with $i,j,k=r,g,b$, respectively. By converting the values of the interfacial tension ratios tested in figure \ref{fig:stability_diagram} to spreading coefficients, our numerically obtained emulsion morphologies are also consistent with their experimental observations.

It is worth noting that we have investigated the optimal value of the coefficient $\beta$ in the additional free energy term introduced in Eq. (\ref{eq:Ea}), varying $\beta$ = 0, 0.0001, 0.001, 0.01, 0.1 and 1.0 for one typical double emulsion morphology at ($\sigma_{gb}/\sigma_{rg}$, $\sigma_{rb}/\sigma_{rg}$) = (1.4, 0.35). The obtained result at $\beta = 0$ (corresponding to the model without the additional term) is shown in figure \ref{fig:stability_diagram} (b). As highlighted by the dashed squares in figure \ref{fig:stability_diagram} (b), two unphysical light blue regions caused by negative $C_1$ appear around the three-phase contact line and lead to incorrect result. The incorrect region is also observed for $\beta = 0.0001$ in figure \ref{fig:stability_diagram} (c). For $\beta$ varying from 0.001 to 0.1, the complete engulfing morphology could be successfully reproduced and invisible difference is observed for different values of $\beta$. However, further increasing $\beta$ to 1.0 induces numerical instability, which indicates that the $\beta$ value cannot be large enough to dominate the double-well potential terms. Meanwhile, for the partial engulfing cases, correct morphologies could be captured even without the additional term, and they are generally unaffected by a small additional term. Based on the above findings, $\beta=0.001$ will be used in subsequent simulations.
	
\section{Results and discussion}
		
\subsection{Previously observed formation regimes and grid independence test}
	
\begin{figure}
	\centering{\includegraphics[scale=0.4]{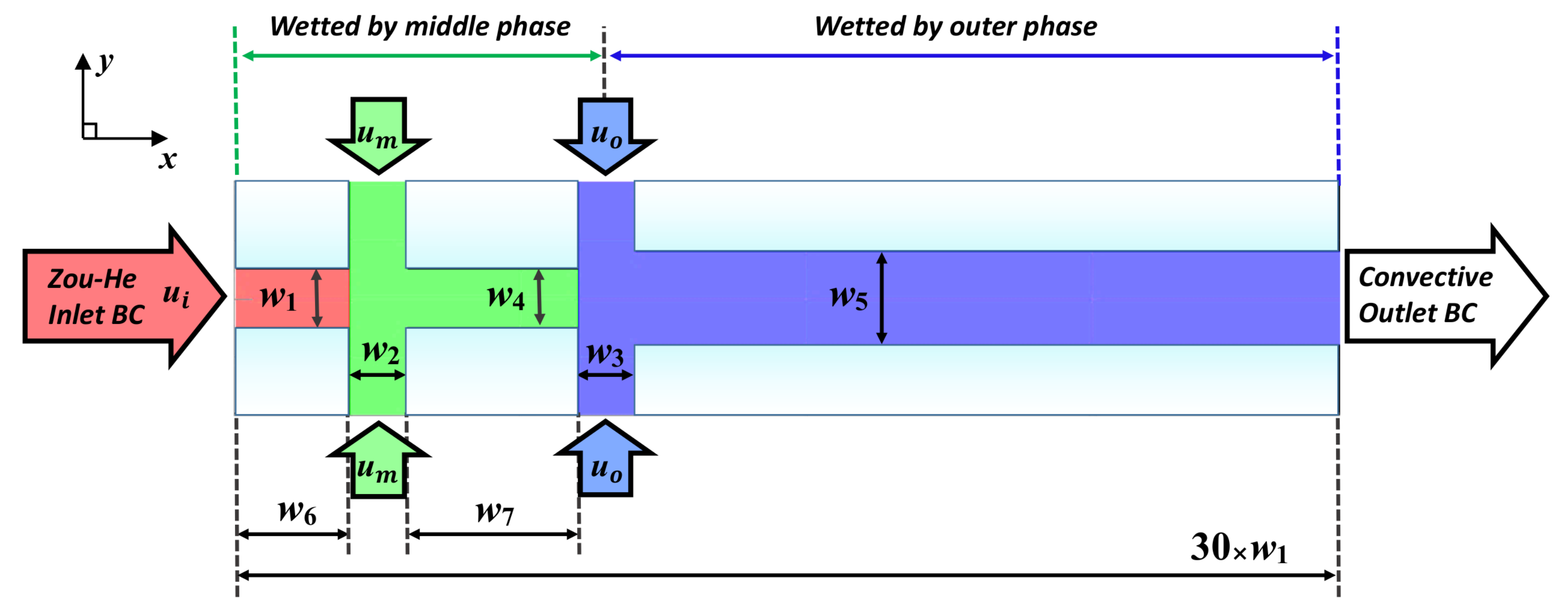}}
	\caption{Illustration of the geometry and boundary settings of the planar hierarchical flow-focusing device in this work.}
	\label{fig:geometry}
\end{figure}

The two-dimensional setup of the hierarchical flow-focusing device is illustrated in figure \ref{fig:geometry}. The inner red fluid is injected through the leftmost inlet with a width of $w_1$, and the middle green and outer blue fluids are injected by two vertical side inlets with widths of $w_2$ and $w_3$, respectively. All the inlet widths are set equal in this section, i.e., $w_1 = w_2 = w_3$. The channel connecting the two side inlets has a width of $w_4 = w_1$, and the main channel width is $w_5 = 1.6w_1$. The length of the first inlet is $w_6 = 2w_1$, and the distance between the two side inlets is $w_7 = 3w_1$. Considering the symmetry of the flow problem in the $y$ direction, only a half of the geometry is simulated and the domain size is $L_x \times L_y = 30w_1 \times 2w_1$. Zou-He velocity inlet boundary condition \citep{Zou1997} is used for all the inlets, and the CBC-AV is applied for the outlet. In addition to the no-slip boundary condition, the wetting boundary condition is also imposed on the solid surfaces, where the first and second junctions are fully wetted by the middle and outer phase fluids, respectively. 

In the following, the subscripts \textit{i}, \textit{m} and \textit{o} are used to denote the inner, middle and outer fluids. The dimensionless parameters that characterize the double emulsion formation process are typically defined as follows \citep{Abate2011,Azarmanesh2016}: the Weber number (the ratio of inertia force to interfacial tension force) of the inner fluid $We_i = \rho_iu_i^2w_1/\sigma_{im}$; Capillary numbers (the ratios of viscous force to interfacial tension force) of the middle and outer fluids $ Ca_m = \eta_m u_m/\sigma_{im}$, $ Ca_o = \eta_o u_o/\sigma_{mo}$; flow rate ratios $Q_i/Q_m = u_i/(2u_m)$, $Q_o/Q_m = (2u_o)/(2u_m) = u_o/u_m$; viscosity ratios $\lambda_{im} = \eta_i/\eta_m$, $\lambda_{om} = \eta_o/\eta_m$; and interfacial tension ratios $\sigma_{io}/\sigma_{im}$ and $\sigma_{mo}/\sigma_{im}$. Here, $u$ is the average inlet velocity. However, in this work, we focus on double emulsion formation behaviors in the limit of small inertia \citep{Nabavi2017M}. As such, it is more appropriate to use $Ca_i = \eta_i u_i/\sigma_{im}$ instead of $We_i$ for the inner fluid. We will change the values of $Ca_i$, $Ca_m$ and $Ca_o$ by adjusting the flow rate of each phase fluid and investigate their roles in formation regime conversions and double emulsion sizes. The viscosity ratios are kept at $\lambda_{im} = \lambda_{om} = 1.24$, and the interfacial tension ratio is fixed at $(\sigma_{mo}/ \sigma_{im}, \sigma_{io} / \sigma_{im})$ = (1.0, 2.2). 

\begin{figure}
	\centering{\includegraphics[scale=0.6]{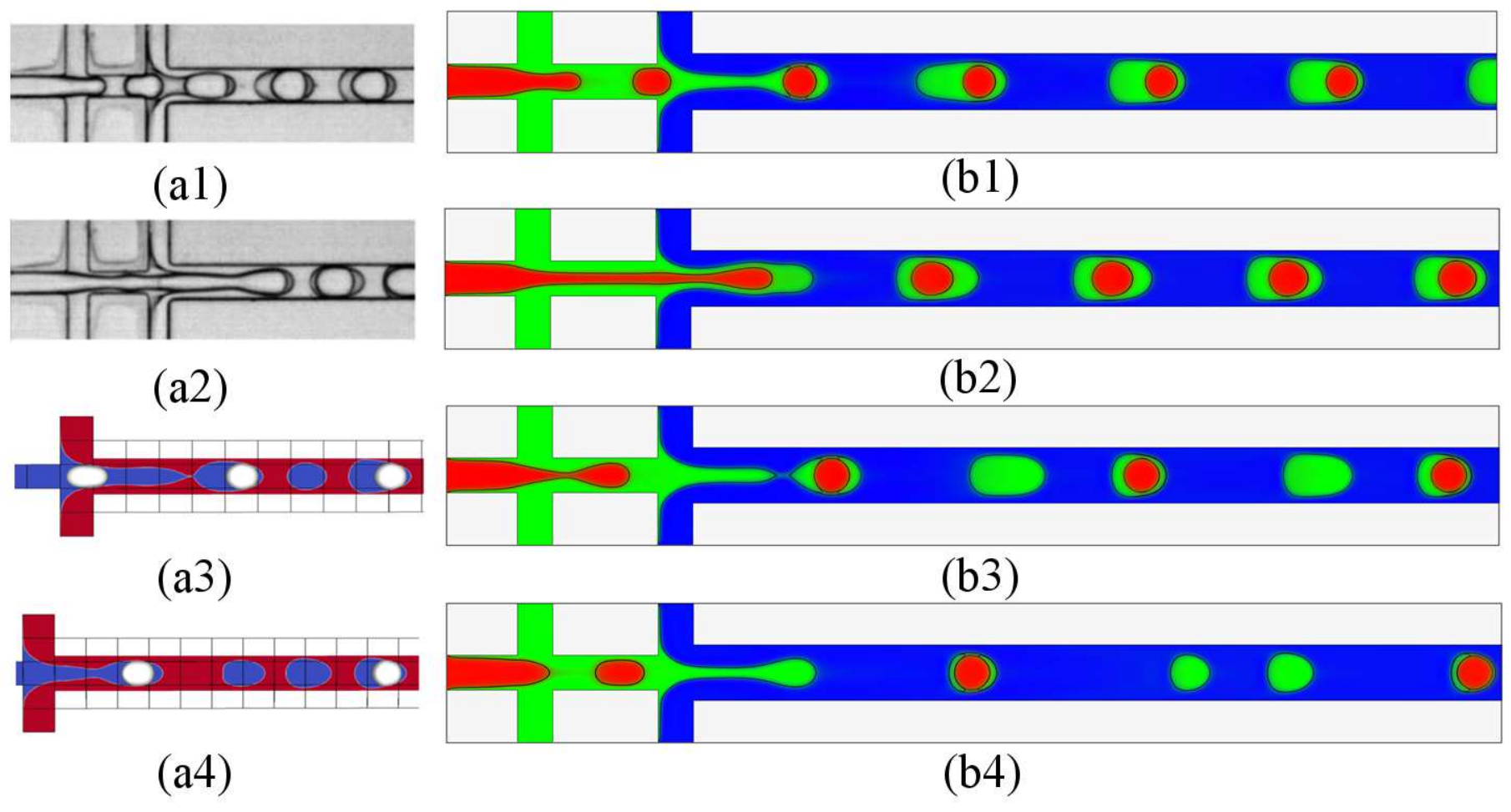}}
	\caption{The present work can qualitatively reproduce common flow regimes previously reported in experimental \citep{Abate2011} (a1-a2) and simulation \citep{Azarmanesh2016} results (a3-a4). Parameters for the present work are (b1) $Ca_i$ = 0.012, $ Ca_m $ = 0.011, and $ Ca_o $ = 0.035; (b2) $Ca_i $ = 0.018, $ Ca_m $ = 0.011, and $ Ca_o $ = 0.035; (b3) $ Ca_i $ = 0.008, $ Ca_m $ = 0.011, and $ Ca_o $ = 0.035; and (b4) $ Ca_i $ = 0.008, $ Ca_m $ = 0.011, and $ Ca_o $ = 0.065. Movies for the cases shown in (b1-b4) are provided online as supplementary materials.}
	\label{fig:basic_regime}
\end{figure}

Four basic flow regimes identified in the double emulsion preparation \citep{Abate2011, Azarmanesh2016} are shown in figure \ref{fig:basic_regime} (a1-a4): (a1) the two-step formation regime, (a2) one-step formation regime, (a3) decussate regime with one empty droplet, and (a4) decussate regime with two empty droplets. Our simulations are able to successfully reproduce all four regimes. Specially, the two-step formation regime is obtained at $Ca_i = 0.012$, $Ca_m = 0.011$ and $Ca_o = 0.035$ (figure \ref{fig:basic_regime} (b1)). With the same $Ca_m$ and $Ca_o$ values, the one-step formation regime is observed by increasing the inner flow rate to $Ca_i = 0.018$ (figure \ref{fig:basic_regime} (b2)), while the decussate regime with one empty middle phase droplet is achieved by decreasing the inner flow rate to $Ca_i = 0.008$ (figure \ref{fig:basic_regime} (b3)). Moreover, if the decussate regime happens at higher $Ca_o$ values, e.g., $Ca_o$ = 0.065 with $Ca_i = 0.005$ and $Ca_m =0.011$, two empty alternate middle phase droplets are found, as shown by figure \ref{fig:basic_regime} (b4). The corresponding $We_i$ values to the $Ca_i$ values used in figure \ref{fig:basic_regime} are generally on the order of $o(10^{-3})$ to $o(10^{-2})$, which are considerably lower than those used in previous studies ($o(1)$) \citep{Abate2011, Azarmanesh2016}. However, we note that similar  two-step and one-step formation behaviors are still obtained.  This suggests that, while $We_i$ can affect the resulting formation regimes, the rich flow behaviors with many different formation regimes are still present in the limit of small inertia \citep{Wu2017single}. Thus, we shall focus here on the limit of small $We_i$ to understand the interplay between viscous and interfacial tension forces. For this reason, it is reasonable to use $Ca_i$ for the inner phase fluid in our study, which also highlights the importance of the flow rate ratios in determining the resulting formation regimes.

A grid independence test is conducted for the two-step formation regime mentioned in figure \ref{fig:basic_regime} (b1). Four different grid resolutions are tested, i.e., $w_1$ = 40, 50, 80 and 100. To make a quantitative comparison, the results from the highest grid resolution ($w_1=100$) is used as a reference. The relative errors ($E_{w_1}=|X_{w_1} - X_{w_1=100}|/X_{w_1=100}$) of the entire emulsion size, pinch-off location and generation period are calculated, and their maximum values are recorded for each grid resolution. The maximum relative errors for $w_1=$ 40, 50 and 80 are $7.18\%$, $4.66\%$, and $0.83\%$, respectively. This suggests that increasing grid resolution from $w_1=50$ to 100 leads to the relative error less than $5\%$, and thus an inlet width of $w_1=50$ is used for the following studies, as a good balance between computational accuracy and cost.
	
\subsection{Effect of flow rates}

\begin{figure}
	\centering{\includegraphics[scale=0.32]{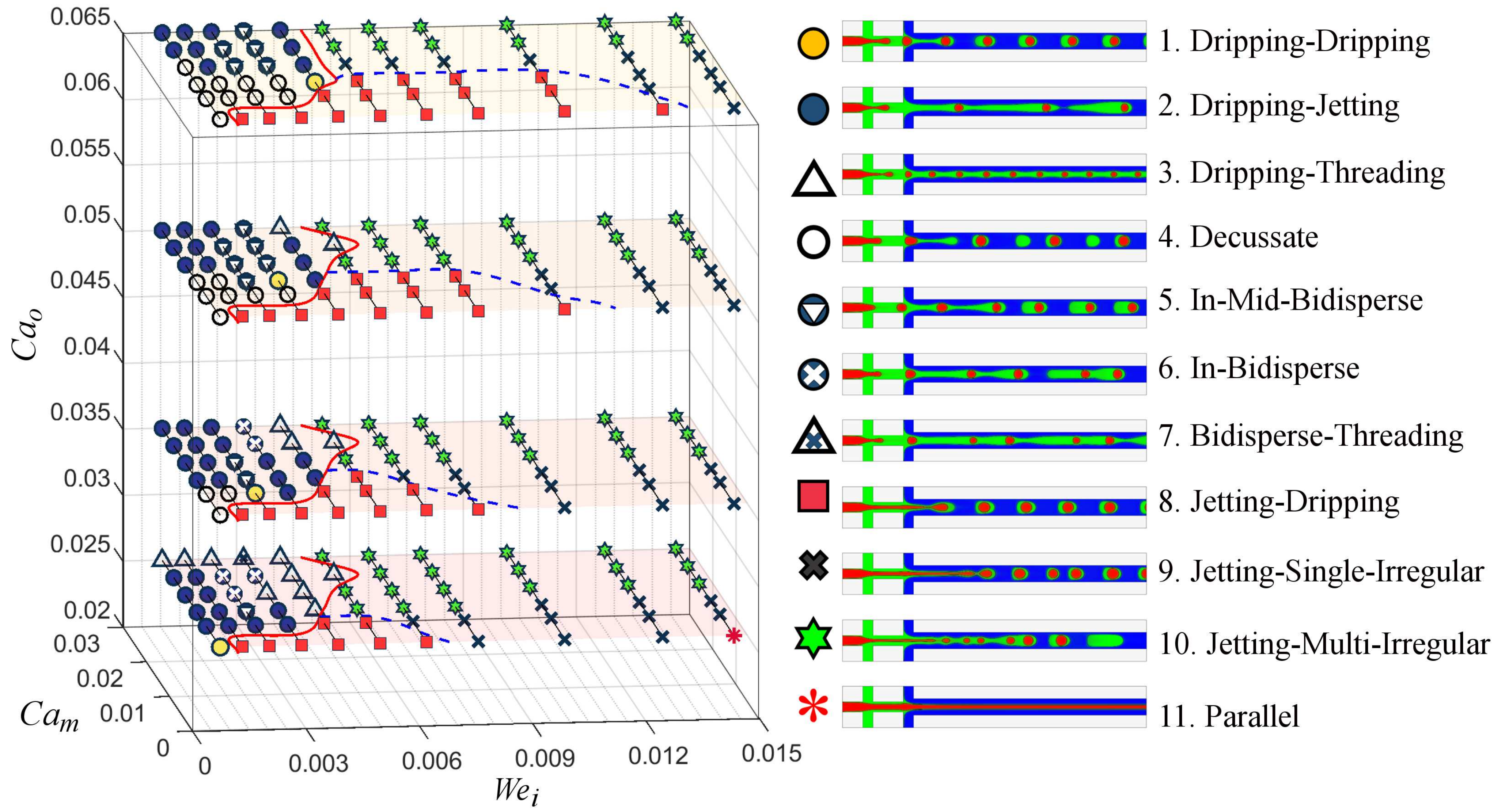}}
	\caption{Flow regimes as a function of $Ca_i$, $Ca_m$ and $Ca_o$.  Each symbol represents a unique formation pattern. Movies are provided online as supplementary materials for Regimes 1-10.}
	\label{fig:phasediagram}
\end{figure}

In the formation of double emulsions, it is known that two-step, one-step and decussate formation regimes can be obtained by varying $Ca_i$, $Ca_m$ and $Ca_o$ values. However, the dependence of each formation regime on $Ca_i$, $ Ca_m $ and $ Ca_o $ values has not been systematically studied, and how the obtained emulsion sizes vary is not very clear. In figure \ref{fig:phasediagram}, a three-dimensional phase diagram is constructed to illustrate how the formation regimes are influenced by $Ca_i$, $ Ca_m $ and $ Ca_o $. The ranges for these influencing parameters are $Ca_i$ = (0.008,  0.01,  0.012,  0.014,  0.016,  0.018,  0.02,  0.022,  0.025,  0.028,  0.03), $Ca_m$ = (0.005,  0.011,  0.015,  0.02,  0.025,  0.03) and $ Ca_o $ = (0.025,  0.035,  0.05,  0.065). It is seen that more formation regimes are obtained besides those reported in figure \ref{fig:basic_regime}. 

To differentiate these regimes, each regime is represented by a unique symbol. The nomenclature for each regime is generally a combination of the breakup modes of the inner and middle phases. To distinguish the dripping and jetting breakup modes, we use the pinch-off location. It is considered as jetting if the distance between the pinch-off location of the inner (or middle) fluid and the downstream edge of the middle (or outer) fluid side inlet is longer than $3w_1$ \citep{Utada2005}. Otherwise, it is categorized as dripping. According to our nomenclature, the periodic two-step and one-step formation regimes shown in figure \ref{fig:basic_regime} (a1) and (a2) are therefore called as Dripping-Dripping (Regime 1) and Jetting-Dripping (Regime 8) regimes in figure \ref{fig:phasediagram}, which distinguishes them from other irregular two-step or one-step formation behaviors. 

\begin{table}
	\begin{center}
		\def~{\hphantom{0}}
		\renewcommand\tabcolsep{7.0pt}
		\begin{tabular}{cll}
			Regime  & Relevant experimental literature   &   Microfluidic device  \\[4pt]
			1-2   & \citet{Abate2011} figure 2   & Two cross-junctions \\
			~     & \citet{Kim2013} figure 3     & Glass capillaries \\
			~     & \citet{Nabavi2015a} figure 2 & Glass capillaries \\[2pt]
			\cline{2-3}
			3     & \citet{Nabavi2017M} figure 4 & Glass capillaries \\
			~     & \citet{Oh2006} figure 6 	 & Assembled flow-focusing device \\[2pt]
			\cline{2-3}
			4     & \citet{Kim2013} figure 2     & Glass capillaries \\[2pt]
			\cline{2-3}
			5     & \citet{Anna2003} figures 2-3 & Single flow-focusing device \\
			~     & \citet{Cubaud2008} Sec V-A 	 & Single cross-junction \\[2pt]
			\cline{2-3}
			6     & \citet{Nabavi2017M} figure 4   & Glass capillaries \\
			~     & \citet{Shang2014} figure 2     & Glass capillaries \\
			~     & \citet{Nisisako2005} figure 2  & Two T-junctions \\[2pt]
			\cline{2-3}
			8     & \citet{Abate2011} figure 2     & Two cross-junctions \\
			~     & \citet{Utada2005} figure 1     & Glass capillaries \\[2pt]	
			\cline{2-3}
			9-10  & \citet{Kim2013} figure 2     & Glass capillaries \\[2pt]
			\cline{2-3}
			11    & \citet{Wu2013} figures 3 	 & Assembled T-Y junctions \\									
		\end{tabular}
		\caption{Relevant experimental literature to the formation regimes shown by figure \ref{fig:phasediagram}}
		\label{tab1}
	\end{center}
\end{table}

To analyze these formation regimes, we classify the formation regimes on each $Ca_o$ plane. Firstly, the formation regimes are divided into two regions by the red solid line according to the breakup mode of the inner phase fluid. The inner phase fluid breaks up in the dripping mode on the left region of the red solid line. All the points in this region are periodic and they could be further subdivided into Regimes 1 to 7. On the right side of the red solid line, the inner phase breaks up in the jetting mode. The right region can be further divided into two subsections by the dashed blue line based on the breakup mode of the middle phase fluid. Below the dashed line, the middle phase fluid breaks up in the dripping mode and we obtain the periodic one-step double emulsion formation regime (Regime 8). Above the dashed line, the middle phase fluid also breaks up in the jetting mode, and the formation behavior loses the periodicity. For instance, the inner and middle phases are pinched off together irregularly (Regime 9), or multiple inner droplets of different sizes are randomly encapsulated in the middle phase droplet (Regime 10). In an extreme case at $Ca_i$ = 0.03, $Ca_m$ = 0.005 and $Ca_o$ = 0.065, parallel layered flow is observed (Regime 11). 

To put the formation regimes obtained in figure \ref{fig:phasediagram} into experimental context, relevant experimental literatures to each formation regime are listed in table \ref{tab1}. Regimes 1, 2, 8, 11 have been reported in planar microfluidic devices, while Regimes 1-4, 8-10 have been observed in capillary microfluidic devices. However, the bidisperse behaviors observed in Regimes 5-7 have only been reported in two-phase experiments so far, some of which are listed to Regime 5 in table \ref{tab1} for reference. A few experiments also present a multiple emulsion formation regime similar to Regime 6 but with two equal-sized inner droplets. Some of these studies are listed to Regime 6 in table \ref{tab1}. In all, figure 6 establishes the connection among different formation regimes. 

In the following sections, we focus on the two-step (Regimes 1-7) and one-step (Regime 8) periodic regions. The effects of flow parameters on the conversion of formation regimes and emulsion sizes will be analyzed in detail to help deepen the understanding of double emulsion formation behaviors.        

\subsubsection{Two-step periodic region}   

In the two-step formation region of figure \ref{fig:phasediagram}, two types of periodic double emulsion formation regimes are observed, i.e., the Dripping-Dripping regime (Regime 1) and the Dripping-Jetting regime (Regime 2). Regime 1 is limited to a small range of governing flow parameters owing to the strict criterion in pinch-off locations. On the other hand, Regime 2 occupies a relatively wider region, and the applicable range of $Ca_m$ for Regime 2 shrinks to higher values as $Ca_o$ increases, due to the appearance of decussate regime at lower $Ca_m$. Moreover, the shape of the red solid lines varies little with $ Ca_o $ over the entire range considered. It indicates that the breakup behavior of the inner phase fluid is mainly determined by $Ca_i$ and $Ca_m$. 

\begin{figure}
	\centering{\includegraphics[scale=0.11]{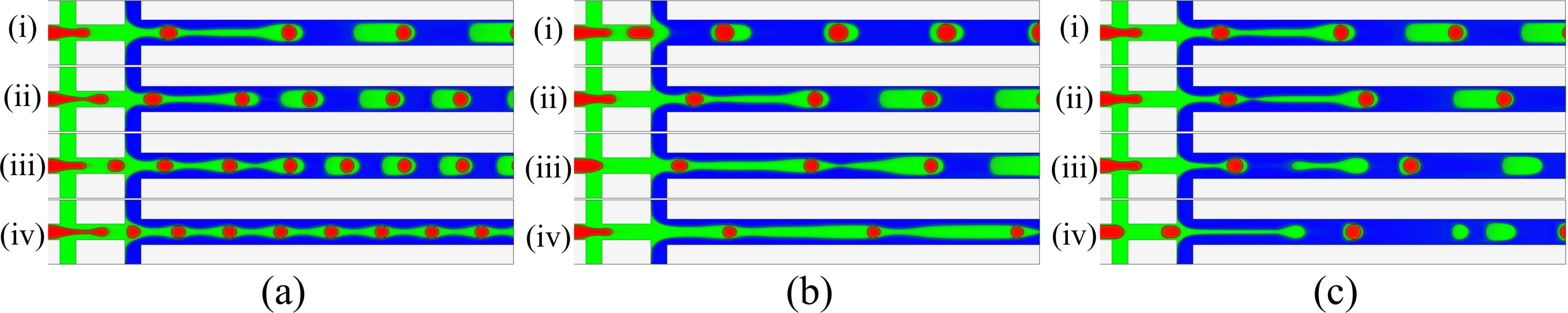}}
	\caption{Dynamics of two-step formation regimes as a function of (a-i)-(a-iv) $Ca_i$ = 0.008, 0.012, 0.014 and 0.016 at $Ca_m$ = 0.015 and $Ca_o$ = 0.025; (b-i)-(b-iv) $Ca_m$ = 0.005, 0.015, 0.02 and 0.03 at $Ca_i$ = 0.008 and $Ca_o$ = 0.025; and (c-i)-(c-iv) $Ca_o$ = 0.025, 0.035, 0.05 and 0.065 at $Ca_i$ = 0.008 and $Ca_m$ = 0.015.} 
	\label{fig:two_step_formation}
\end{figure}

To clarify the effects of $Ca_i$, $Ca_m$ and $Ca_o$ on the two-step formation regimes, we illustrate the typical formation behaviors as a function of the $Ca_i$, $Ca_m$ and $Ca_o$, respectively in figure \ref{fig:two_step_formation}. The parameters are (a) $Ca_i$ = 0.008, 0.012, 0.014 and 0.016 at $Ca_m$ = 0.015, $Ca_o$ = 0.025; (b) $Ca_m$ = 0.005, 0.015, 0.02 and 0.03 at $Ca_i$ = 0.008, $Ca_o$ = 0.025; and (c) $Ca_o$ = 0.025, 0.035, 0.05 and 0.065 at $Ca_i$ = 0.008, $Ca_m$ = 0.015. Based on figure \ref{fig:two_step_formation}, we will discuss the size variations of double emulsion generated in the two-step formation regime. Moreover, new insights on other typical formation regimes obtained in the ternary system will also be discussed.

\smallskip
{\flushleft
\paragraph{(1) Size variations of double emulsions generated in the two-step regime}
}
\smallskip

For the typical two-step formation regime shown in figure \ref{fig:two_step_formation}, the dripping/squeezing breakup mode of the inner phase fluid in the ternary system is similar to that happens in a binary system. It is generally attributed to the action of the leading viscous force and the squeezing effect that overcome the interfacial tension force \citep{Fu2012, Cubaud2008, Yuwei2019}. For the breakup of the middle phase fluid, it is subject to both the viscous force of the outer phase fluid and the resulting flow from the generated inner droplets. The expansion of the middle phase front in the main channel also leads to accumulated upstream pressure in the outer phase fluid. All these factors assist in the breakup of the middle phase fluid.

Figure \ref{fig:two_step_formation} (a-i)-(a-III) shows the effect of $Ca_i$ on the two-step double emulsion formation behaviors. With increasing $Ca_i$, the inner droplet size decreases and its formation frequency increases. This trend has also been numerically observed by \citet{Fu2016} for double emulsions generated at high flow rates of the middle phase fluid in a two dimensional simplified co-axial device. The increased formation frequency of the inner phase droplet shortens the time to accumulate the upstream pressure in the outer phase fluid and actuates the pinch-off of the middle phase front. Thus, the size of the middle layer of the entire double emulsion also decreases. 

The effect of $Ca_m$ on two-step formaiton behaviors are given in figure \ref{fig:two_step_formation} (b). From figure \ref{fig:two_step_formation} (b-i) to (b-iii), it is noticed that the inner droplet size decreases while the middle part of the double emulsion increases. As the intermediate layer, the middle phase fluid has dual effects. With increasing $Ca_m$, the increased viscous force of the middle phase fluid exerted on the inner phase fluid leads to the size reduction of the inner droplet. Meanwhile, the increased middle flow rate decreases its velocity difference to that of the outer phase, which effectively lowers the outer shear stress and extends the time for the middle part of the double emulsion to grow larger. Our results show that the increase in the middle part size is usually more significant than the decrease in the inner droplet size. Thus, the entire double emulsion size increases. 
	
Figure \ref{fig:two_step_formation} (c) displays the effects of $Ca_o$ on double emulsion formation behaviors. As seen in figure \ref{fig:two_step_formation}(c-i)-(c-ii), increasing $Ca_o$ does not change the double emulsion size in any obvious way before the formation regime changes, but the distance between the generated double emulsions gets larger due to the increased outer flow rate. Further increasing $Ca_o$ to 0.05, decussate regime with one alternate empty droplet appears and the shell thickness of the double emulsion is greatly reduced. When $Ca_o$ reaches 0.065, double emulsions with two alternate empty droplets are captured.  

\smallskip
{\flushleft
\paragraph{(2) Dripping-threading regime}
}
\smallskip

For the periodic two-step region shown in figure \ref{fig:phasediagram}, increasing $Ca_i$ or $Ca_m$ would both lead to the dripping-threading regime, where the inner droplet is produced periodically in the continuous middle phase thread. Two examples are given in figure \ref{fig:two_step_formation} (a-iv) and (b-iv). We would like to compare the differences between the dripping-threading morphologies obtained by adjusting $Ca_i$ and $Ca_m$, respectively. The inner droplets are produced in small sizes in both cases, while the formation frequency is higher at large $Ca_i$ than that obtained at large $Ca_m$. As a result, the capillary perturbations on the middle phase fluid in case (a-iv) is counteracted by the high formation frequency of the inner phase droplets and the obtained dripping-threading regime is very stable. \citet{Nabavi2017M} experimentally reported a similar stable dripping-threading regime also by increasing the inner flow rate in a capillary device. Unlike in case (a-iv), it is seen in (b-iv) that some necking regions develop at the middle phase fluid as it flows downstream, which would possibly lead to the middle phase breakup somewhere more downstream. Such unstable regimes as shown in figure \ref{fig:two_step_formation} (b-iii)-(b-iv) remind us of the varicose shape reported in binary experiments \citep{Cubaud2008}. The narrow main channel limits the expansion of the middle phase front to form an emulsion, and the following embryonic emulsion shape begins to grow before the front one is pinched off. 

In view of applications, the compound structure generated in the Dripping-Threading regime is capable of producing bundles of microcapsules that are promising for storing, handling and arrayed assay of small volumes \citep{Oh2006}. To remove the Dripping-Threading regime, we can increase $Ca_o$ to produce regular double emulsions.

\smallskip
{\flushleft
\paragraph{(3) Bidisperse formation regime}
}
\smallskip

\begin{figure}
	\centering{\includegraphics[scale=0.6]{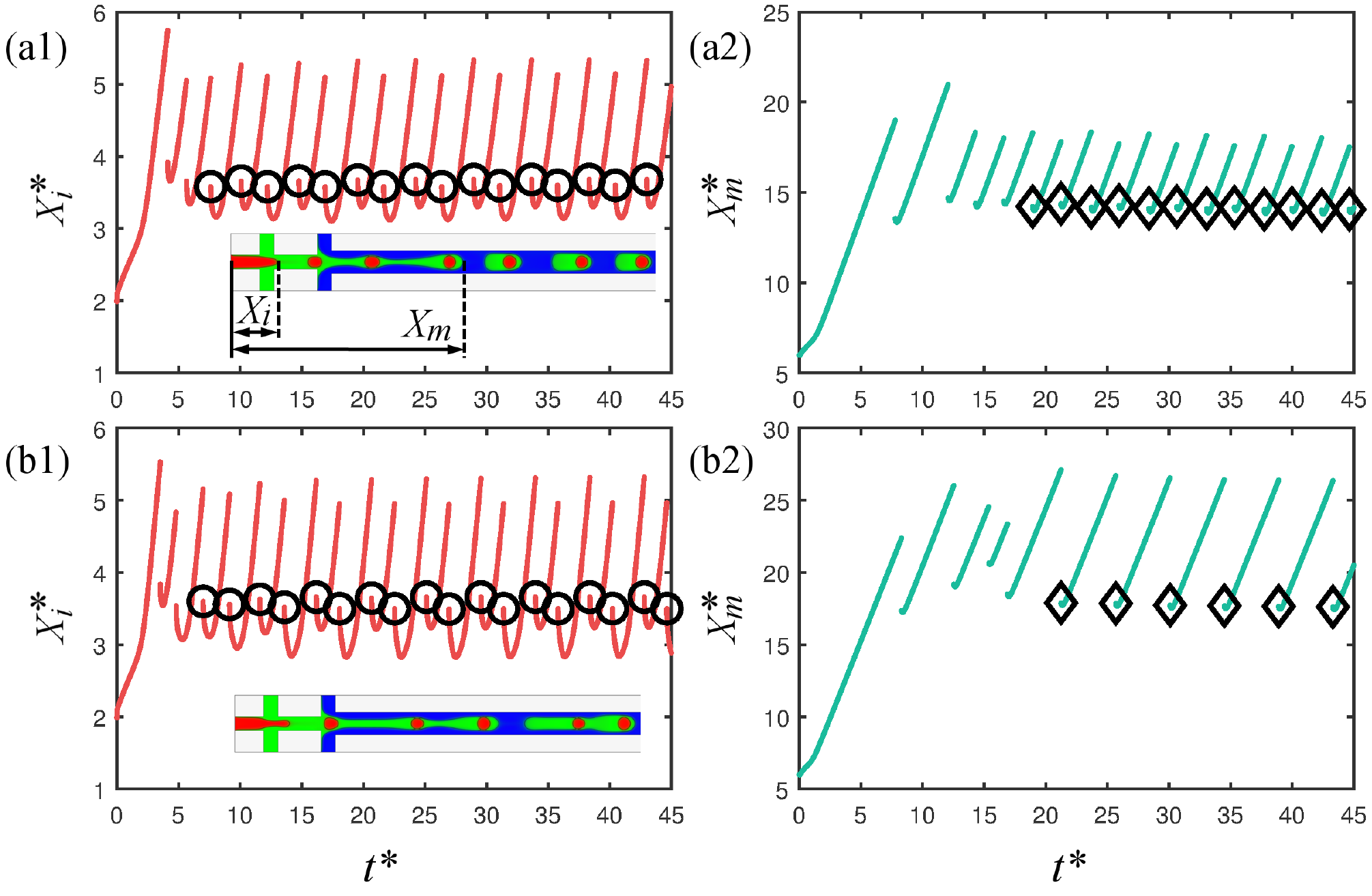}}
	\caption{Temporal evolutions of the thread tip locations of the inner ($X^*_i$) and middle ($X^*_m$) phases obtained for (a1, a2): $Ca_i$ = 0.012, $Ca_m$ = 0.015 and $Ca_o$ = 0.025, and (b1, b2) $Ca_i$ = 0.012, $Ca_m$ = 0.02 and $Ca_o$ = 0.025. The time and location are normalized using $t^* = t(u_m)_{max}/w_1$ and $ X^*_{i,m} = X_{i,m}/w_1$, where $(u_m)_{max}$ = 0.0015 is the maximum flow rate of the middle phase used in the current study. The superimposed empty round circles in (a1, b1) and diamond symbols in (a2, b2) mark the periodic points that correspond to each pinch-off moment and location of the inner and middle phases, respectively. The inset snapshots in (a1) and (b1) show the corresponding flow behaviors in each case.}
	\label{fig:pinch_off_bidisperse}
\end{figure}

In figure \ref{fig:two_step_formation} (a-ii), size variations are observed in the generated double emulsion sequence in the main channel: a smaller double emulsion is followed by a larger one, and this pattern repeats itself. To reveal the periodicity of this behavior, the temporal evolution of the inner and middle thread tip locations are traced as denoted by $X_i$ and $X_m$ in figure \ref{fig:pinch_off_bidisperse} (a1) and (a2), respectively. The time $t$ and locations $X_{i,m}$ are normalized using $ t^* = t(u_m)_{max}/w_1$ and $ X^*_{i,m} = X_{i,m}/w_1$, where $(u_m)_{max}$ = 0.0015 is the maximum flow rate of the middle phase fluid used in the current study. After the double emulsions are produced regularly, the points corresponding to the pinch-off moment and location in each formation period are marked by the superimposed round circles for the inner phase fluid in figure \ref{fig:pinch_off_bidisperse} (a1) and diamonds for the middle phase fluid in figure \ref{fig:pinch_off_bidisperse} (a2). Clearly, periodic fluctuations in pinch-off locations and formation periods are observed in both the inner and middle phases between every two neighboring droplets, which is consistent with the variation in emulsion sizes observed in figure \ref{fig:two_step_formation} (a-ii). This flow pattern is named as the In-Mid-Bidisperse regime (Regime 5). 

Regime 5 is frequently observed for $Ca_i$ between 0.012 and 0.014 with $Ca_m$ approximately from 0.015 to 0.03 on each $Ca_o$ plane. The earlier occurrence time of the inner phase bidispersity observed in figure \ref{fig:pinch_off_bidisperse} (a1) and (a2) suggests that such bidisperse behaviors mostly originate from the inner phase fluid and then propagate to the middle phase fluid. Since the breakup mode of the inner phase fluid is rarely affected by $Ca_o$, the reason for the inner phase bidispersity should be similar to that in a binary system. It is normally attributed to the oscillations in the amount of residual liquid on the entrance side after the previous droplet is pinched off \citep{Coullet2005,Utada2007,Garstecki2005Nonlinear}.

Noteworthy, the influence of inner phase bidispersity brings richer dynamics in the present ternary system depending on $Ca_m$ and $Ca_o$ values. For cases like the one shown in figure \ref{fig:two_step_formation} (a-ii), the middle phase fluid is easily to be pinched off and it follows the bidisperse breakup frequencies of the inner phase fluid (Regime 5). However, for a flow condition with a high $Ca_m$ and a low $Ca_o$, the thicker middle phase fluid could extend its pinch-off time and engulf every two inner phase droplets inside, as shown by one typical case at $Ca_i = 0.012$, $Ca_m = 0.02$ and $Ca_o = 0.025$ in figure \ref{fig:pinch_off_bidisperse} (b1)-(b2). As such, the variation in formation frequency only happens in the inner phase fluid (figure \ref{fig:pinch_off_bidisperse} (b1)), but not in the middle phase fluid (figure \ref{fig:pinch_off_bidisperse} (b2)). It is named as the In-Bidisperse regime (Regime 6). In addition, even if the middle phase fluid forms a continuous thread, e.g., at $Ca_i = 0.014$, $Ca_m = 0.03$ and $Ca_o = 0.025$, the inner bidisperse behavior could still happen, and it is named as the Bidisperse-Threading regime (Regime 7). 

\smallskip
{\flushleft
\paragraph{(4) Decussate regime with two empty droplets}
}
\smallskip

\begin{figure}
	\centering{\includegraphics[scale=0.5]{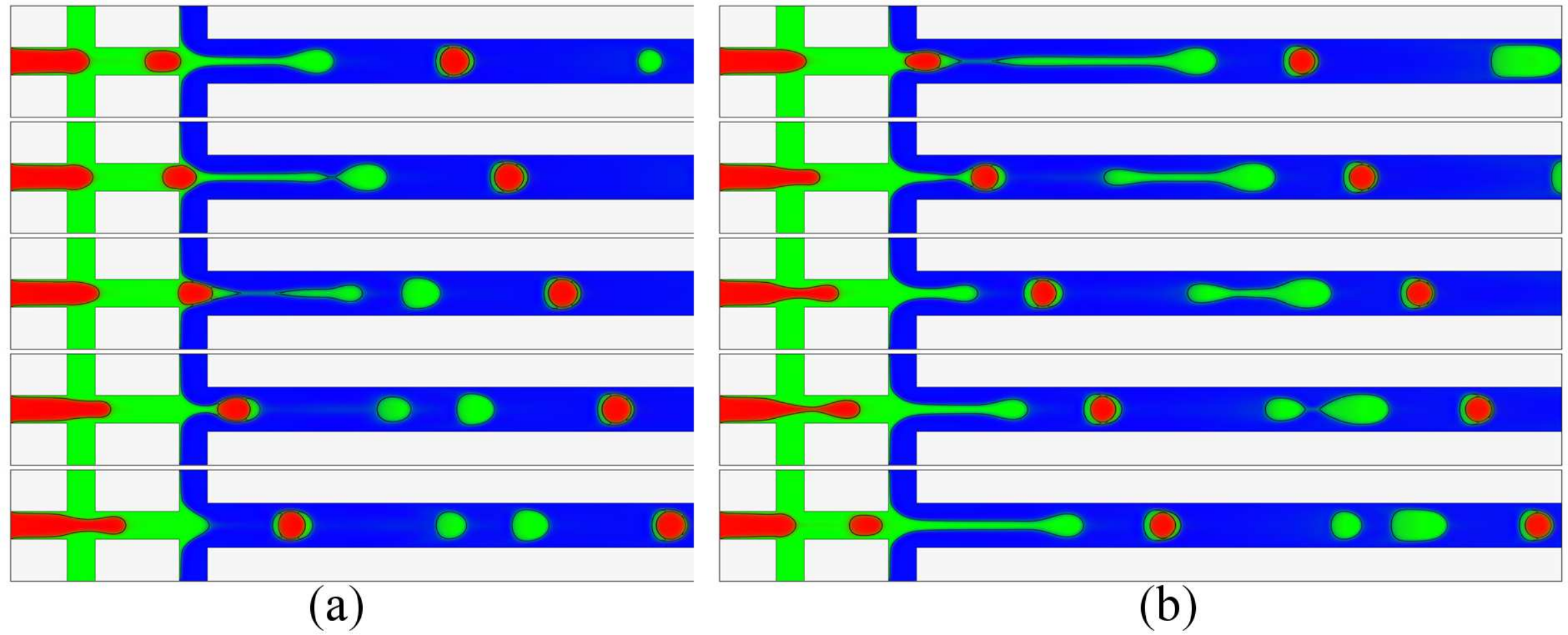}}
	\caption{Decussate regimes with two empty droplets: (a) $Ca_i$ = 0.008, $Ca_m$ = 0.015, and $Ca_o$ = 0.065; and (b) $Ca_i$ = 0.008, $ Ca_m $ = 0.011, and $ Ca_o $ = 0.065.}
	\label{fig:decussate_detail}
\end{figure}

Decussate regimes occupy a substantial proportion in the two-step formation region of figure \ref{fig:phasediagram}. Among them, decussate regimes with one alternate empty droplet is commonly observed while decussate regimes with two empty droplets mainly happen at high $Ca_o$ values. Figure \ref{fig:two_step_formation} (c-iv) gives one example of the decussate regime with two empty droplets, and the formation process of the two empty droplets is shown in figure \ref{fig:decussate_detail} (a). It is seen that a long section of the middle phase fluid is pinched off entirely, and then it breaks up into two daughter droplets during the retraction process of the stretched structure when flowing downstream. \citet{Azarmanesh2016} numerically reported another type of formation process for decusste regime with two empty droplets, where the two empty droplets are produced one by one. Our results show that by lowering $Ca_m$ of the case shown in figure \ref{fig:decussate_detail} (a) to 0.011 in figure \ref{fig:decussate_detail} (b), the formation process reported by \citet{Azarmanesh2016} is reproduced in our work. 

Comparing figure \ref{fig:decussate_detail} (a) and (b), the only difference lies in $Ca_m$. A lower $Ca_m$ signifies a higher velocity difference between the middle and the outer phases, which leads to a stronger viscous force exerted on the middle phase fluid and contributes to the early pinch-off of the middle phase front around the bulb neck. With regard to figure \ref{fig:decussate_detail} (a) at a higher $Ca_m$, the middle phase front is not pinched off until the entrance of the inner droplet that prevents the continuous injection of the middle phase fluid to its thread tip.

Decussate regimes are also of practical significance. For instance, if the downstream channel is connected to an expansion channel, the empty droplet can catch up with the double emulsion droplet ahead and merge to form a large double emulsion with thicker middle layer \citep{Azarmanesh2016}. Moreover, the empty droplet and the double emulsion droplet can be viewed as two distinct inner components to produce more complex functional multiple emulsions \citep{Nisisako2005}.

\subsubsection{One-step periodic region}

Even though both the two-step (Regimes 1-2) and one-step (Regime 8) formation regimes can be used for producing double emulsions, the one-step regime is advantageous over the two-step regime in several aspects, such as better robustness in wetting conditions, producing double emulsions with thinner middle layer and emulsifying non-Newtonian fluids \citep{Abate2011}. Moreover, from the point of view of the applicable parameter ranges shown in figure \ref{fig:phasediagram}, a wide range of one-step formation points distributed continuously, different from the distribution character of the periodic two-step double emulsion formation regimes with interference from other flow regimes. Therefore, the periodic one-step double emulsion region has more statistical significance over the periodic two-step region. It enables us to investigate the one-step formation mechanism more quantitatively and construct possible scaling laws for the double emulsion sizes.
\begin{figure}
	\centering{\includegraphics[scale=0.1]{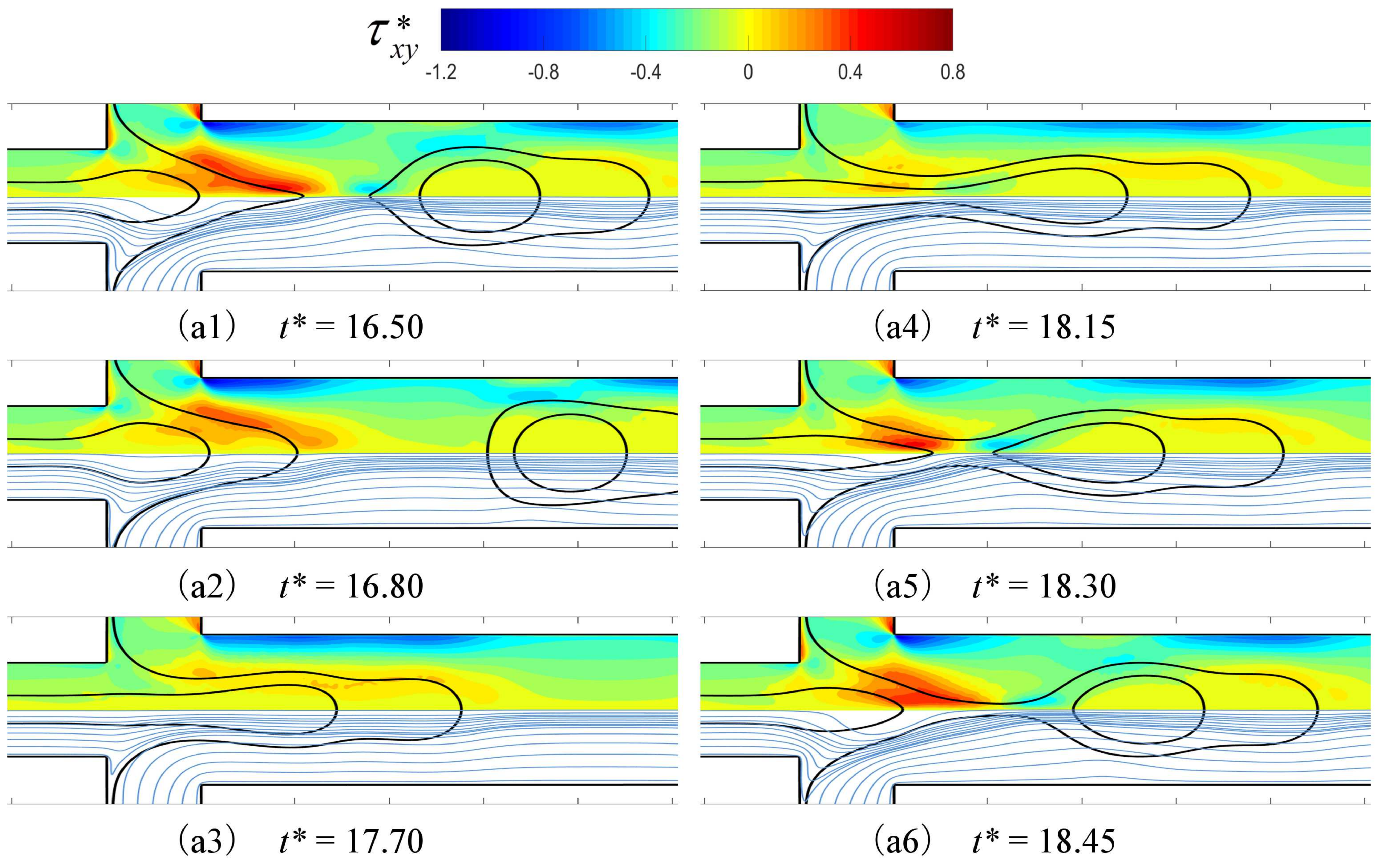}}
	\caption{Time evolution of the interface dynamics at $Ca_i$ = 0.018, $Ca_m$ = 0.011 and $Ca_o$ = 0.05. In each subfigure, the interface shapes are denoted by the solid lines. The distribution of the normalized viscous force component $\tau^*_{xy} = \tau_{xy}w_1/\sigma_{im}$ is shown in the upper part, while the streamlines are shown in the lower part. The time is normalized using $t^* = t(u_m)_{max}/w_1$.}
	\label{fig:evolution}
\end{figure}

In figure \ref{fig:phasediagram}, the applicable range of $Ca_i$ for Regime 8 increases with $ Ca_o $ and decreases with $Ca_m$. The latter trend is consistent with the experimental observation of \citet{Kim2013} in the study of periodic one-step double emulsion using a capillary device. To better understand the one-step double emulsion formation process, the typical temporal evolution of the interface dynamics at $Ca_i$ = 0.018, $Ca_m$ = 0.011 and $Ca_o$ = 0.05 is shown in figure \ref{fig:evolution}. In each sub-figure, the interface shapes are depicted by the solid lines. The leading viscous force component is displayed in the upper part, i.e., $\tau_{xy} = \eta(\p u_y/ \p x + \p u_x/ \p y )$ for a two-dimensional system, and it is normalized using $\tau^*_{xy} = \tau_{xy}w_1/\sigma_{im}$. The streamlines are shown in the lower part. Figure \ref{fig:evolution} (a1) corresponds to the moment just after a previous double emulsion is pinched off, where a strong shear stress region is activated to resist the retraction of the highly deformed middle-outer interface. During the evolution from figure \ref{fig:evolution} (a1) to (a3), the middle phase thread tip approximately recovers to a semicircular shape under the effect of interfacial tension \citep{Utada2007,Fu2012}. In the meantime, the highest shear stress is lowered, and a more evenly distributed high shear stress region is formed along the inner-middle interface. The inflation of the compound inner and middle thread tip partially blocks the inflow of the outer phase fluid. Then, the outer fluid squeezes back the expanded compound thread tip and stretches it downstream. An obvious neck region is formed in figure \ref{fig:evolution} (a4) and it keeps shrinking until the pinch-off happens in the inner phase fluid as shown in figure \ref{fig:evolution} (a5). It is seen that a higher positive and a lower negative shear stress regions are induced immediately near the newly pinched inner thread tip and the generated inner droplet, respectively. The weakly connected middle thin thread is pinched off just after the configuration in figure \ref{fig:evolution} (a6).

Based on the analysis of figure \ref{fig:evolution}, the double emulsion formation process in the one-step regime can be approximately viewed as the sum of a partial blocking period and a squeezing period, which is analogous to that of a single droplet formation process in squeezing or dripping regime in binary flow-focusing systems \citep{Cubaud2008,Liu2011,Fu2012}. Thus, the scaling laws proposed in binary flow-focusing systems could be used as the basis for the construction of scalling laws of the present ternary system. To complete the scaling law involving the ternary flow characters, we first analyze the effects of governing flow parameter on the double emulsion sizes.

\smallskip
{\flushleft
\paragraph{(1) Size variations of double emulsions generated in the one-step regime}
}
\smallskip

\begin{figure}
	\centering{\includegraphics[scale=0.3]{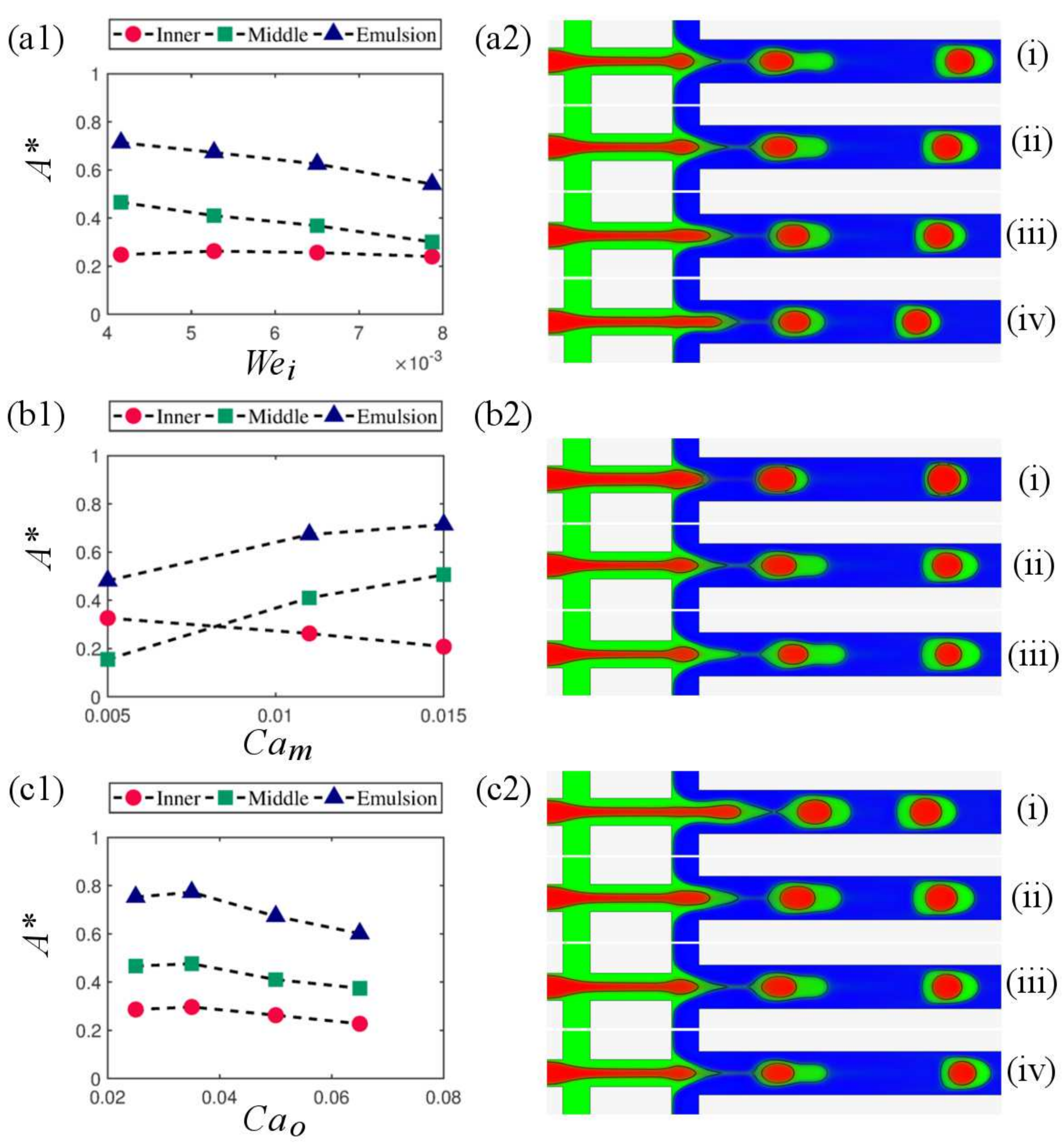}}
	\caption{Inner part, middle part and the entire double emulsion size variations in the periodic one-step formation regime (Regime 8) by changing: (a1) $Ca_i$ = 0.016, 0.018, 0.02 and 0.022 at $Ca_m$ = 0.011 and $Ca_o$ = 0.05; (b1) $Ca_m$ = 0.005, 0.011 and 0.015 at $Ca_i$ = 0.018 and $Ca_o$ = 0.05; and (c1) $Ca_o$ = 0.025, 0.035, 0.05 and 0.065 at $Ca_i$ = 0.018 and $Ca_m$ = 0.011. The snapshots shown in (a2-i)-(a2-iv), (b2-i)-(b2-iii) and (c2-i)-(c2-iv) correspond to the flow conditions mentioned in (a1)-(c1).}
	\label{fig:one_step_formation}
\end{figure}

As shown in figure \ref{fig:one_step_formation} (a1), the effect of $Ca_i$ is investigated for $Ca_i$ = 0.016, 0.018, 0.02 and 0.022 at $Ca_m$ = 0.011 and $Ca_o$ = 0.05. The areas of the entire double emulsion $A_{emulsion}$, the inner part $A_{inner}$, and the middle part $A_{middle}$ are measured after the double emulsion is produced periodically. The area quantities are normalized using $A^* = A/(\upi w_1^2)$. Compared to the effect of increasing $Ca_i$ in the two-step double emulsion formation regime given in figure \ref{fig:two_step_formation} (a-i)-(a-iii), the inner droplet size decreases in the two-step formation regime, while it varies little except for an initial minor increase in the one-step formation regime. Since the inner droplet size varies little, the time needed for the inner phase fluid to breakup is shortened with the increase of $Ca_i$ \citep{Utada2007}. This leads to the size reduction in the middle part and the entire double emulsion size, which is qualitatively similar to the effect of $Ca_i$ observed in two-step formation regimes. By investigating all the periodic one-step data shown in figure \ref{fig:phasediagram}, the variations in double emulsion sizes caused by $Ca_i$ are qualitatively the same for other $Ca_m$ and $Ca_o$ conditions.

The effect of $Ca_m$ on the size of each component and the corresponding snapshots are illustrated in figure \ref{fig:one_step_formation} (b1) and (b2) at $Ca_m$ = 0.005, 0.011 and 0.015 with $Ca_i$ = 0.018 and $Ca_o$ = 0.05. As $Ca_m$ increases, the inner droplet size decreases and the middle part increases. The middle part size always increases faster than the decrease of the inner droplet size. Thus, the entire double emulsion size increases monotonously with $Ca_m$. These trends qualitatively agree with the size variation characters obtained in the two-step regimes as shown in figure \ref{fig:two_step_formation} (b-i)-(b-iii). It indicates the same effects of $Ca_m$ on both formation regimes. We further verify that varying $Ca_i$ and $Ca_o$ conditions in figure \ref{fig:phasediagram} does not change the effects of $Ca_m$.

We have learned the effects of $Ca_o$ on two-step formation regimes in figure \ref{fig:two_step_formation} (c): the inner droplet size is almost independent of $Ca_o$, but the breakup frequency of the middle phase increases with increasing $Ca_o$, which could further lead to the decussate regime. However, a different effect of $Ca_o$ is expected in the one-step formation regime since the inner and middle phase fluids are emulsified simultaneously. In figure \ref{fig:one_step_formation} (c1) and (c2), $Ca_o$ is increased from 0.025, 0.035, 0.05 to 0.065 at $Ca_i$ = 0.018 and $Ca_m$ = 0.011. As $Ca_o$ increases, identical variation trends occur to the inner part, middle part and the entire double emulsion sizes: the sizes consistently increase slightly at the very beginning and then decrease monotonously. For other $Ca_i$ and $Ca_m$ values investigated in figure \ref{fig:phasediagram}, the initial increase in sizes is not common with increasing $Ca_o$, but the decreasing trend is always obtained due to the enhanced viscous force at larger $Ca_o$. Therefore, for the purpose of constructing the scaling law on the double emulsion sizes, the occasional increasing trend is neglected, and we will assume the size has a decreasing trend with increasing $Ca_o$.

\smallskip
{\flushleft
\paragraph{(2) Scaling laws for character sizes of the double emulsion}
}
\smallskip

To construct a phenomenological scaling law for the size of double emulsion produced in the one-step regime, we take inspiration from the scaling laws developed for the size of single droplet generated in squeezing regime within a single cross junction. Several researchers have contributed to the development of droplet size scaling laws in such binary systems \citep{Garstecki2006,Tan2008, Christopher2008, Liu2011}. Specifically, \citet{Liu2011} developed a scaling law for the length $L_p$ of the obtained plug shape droplet, which is given by
\begin{equation}
\frac{L_p}{w_1} =(\tilde{\epsilon} + \tilde{\gamma} \frac{Q_{dispersed}}{Q_{continuous}}) Ca_o^{\tilde{m}}, 
\label{eq:pluglength}
\end{equation}  
where the plug length is normalized by the inlet width $w_1$, and $\tilde{\epsilon}$, $\tilde{\gamma}$ and $\tilde{m}$ are fitting parameters. $Q_{dispersed}/Q_{continuous}$ is the flow rate ratio between the dispersed droplet phase and the continuous carrier phase. 

In Eq. (\ref{eq:pluglength}), the contributions of the blocking and squeezing processes for the size of the obtained droplet are described by the first and second terms in the bracket, respectively. It also includes the power-law dependence of the droplet size on the outer phase capillary number as pointed out by \citet{Christopher2008}. Moreover, the work of \citet{Liu2011} showed that for droplet produced at different width or height conditions, the fitting parameters will be affected, but the variation of droplet size still obeys the generalized expression of Eq. (\ref{eq:pluglength}). The good agreement with available results justifies the validity of this scaling law \citep{Liu2011}, and it is therefore used as the basis to construct a size scaling law for the double emulsion.

Besides the similarities, we would like to highlight the differences between the binary and ternary systems so as to extend Eq. (\ref{eq:pluglength}) for the ternary system. Firstly, the droplet length is only suitable for plug shape droplets whose diameter is wider than the channel width \citep{Garstecki2006, Liu2011}. The volume values are more general quantities for the ellipsoid-like double emulsions \citep{Steegmans2009, Chang2009, Fu2016}. The volume quantities are also more convenient to measure the size of each part of the double emulsion than the length quantities. Hence, the areas of each part of the double emulsion are monitored as discussed in the above subsection. The equivalent radius of the double emulsion area defined by $R_{emulsion} = \sqrt{A_{emulsion}/\upi}$ is used as the dependant variable of the scaling law. Secondly, the dispersed phase in the one-step formation regime is made up of both the inner and middle phases for the continuous outer phase fluid. The independent control over the inner and middle flow properties make the flow behaviors more complex. Based on the analysis of flow parameter effects shown in figure \ref{fig:one_step_formation}, $Q_{dispersed}/Q_{continuous}$ in Eq. (\ref{eq:pluglength}) is replaced by $Q_m/Q_i$ to incorporate the positive effect of $Ca_m$ and the negative effect of $Ca_i$ on the entire double emulsion size.

\begin{figure}
	\centering{\includegraphics[scale=0.5]{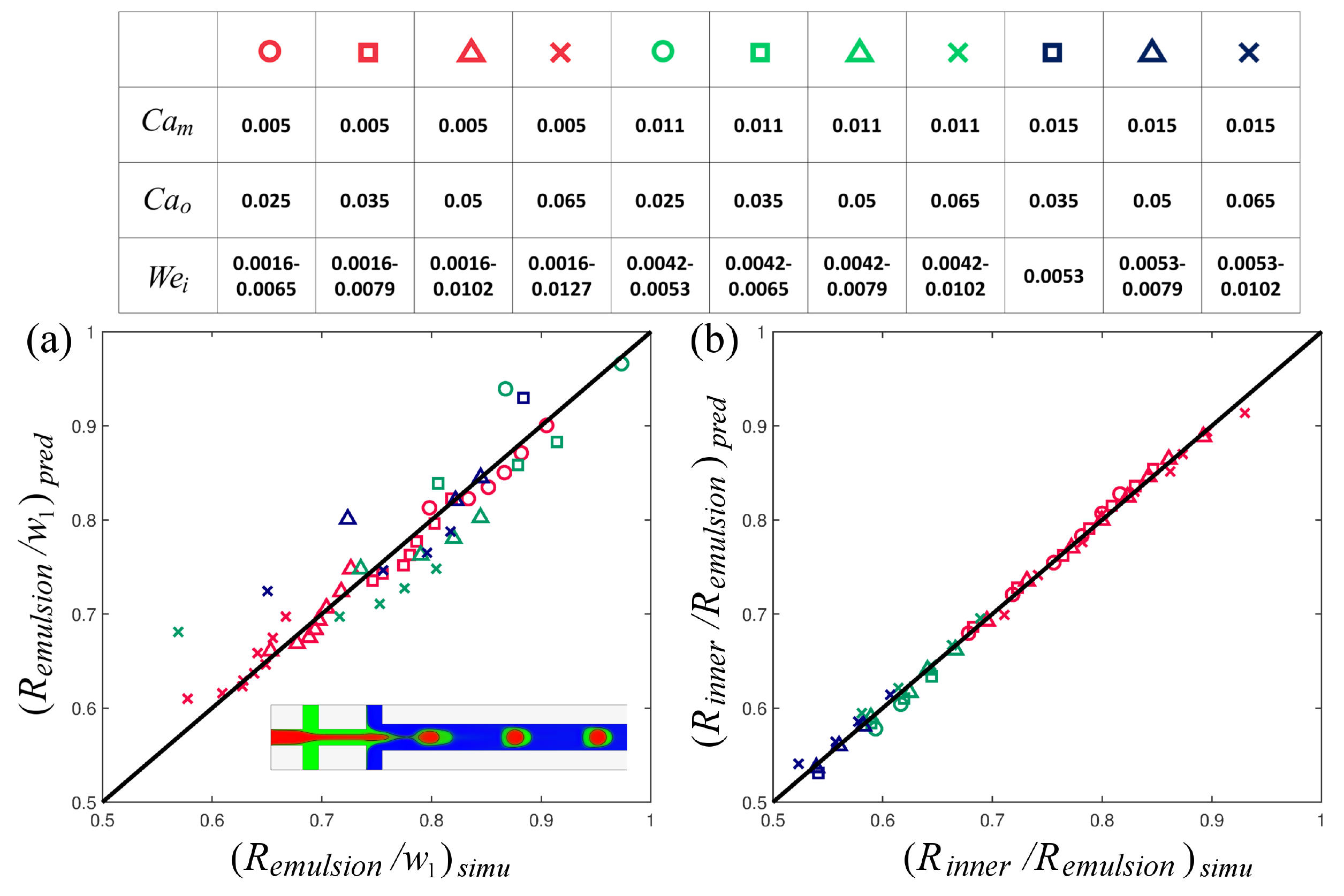}}
	\caption{The parity plots of (a) the normalized entire double emulsion radius $(R_{emulsion}/w_1)_{pred}$ computed from Eq. (\ref{eq:EW}) and the simulated values of $(R_{emulsion}/w_1)_{simu}$; and (b) the radius ratio of the inner part to the entire double emulsion $(R_{inner}/R_{emulsion})_{pred}$ computed from Eq. (\ref{eq:IE}) and the simulated values of $(R_{inner}/R_{emulsion})_{simu}$. The points in both plots are based on all periodic one-step flow conditions (Regime 8) obtained in figure \ref{fig:phasediagram}. The legend table shows that the flow conditions of all feasible $Ca_i$ at each $Ca_m$ and $Ca_o$ combination are represented by symbols of the same type, and the values of $Ca_m$ and $Ca_o$ are differentiated through the symbol colors and shapes, respectively. The inset in sub-figure (a) shows the snapshot of one typical periodic one-step formation regime at $Ca_i$ = 0.02, $Ca_m$ = 0.011 and $Ca_o$ = 0.05.}
	\label{fig:EW_IE}
\end{figure}

Thus, the scaling law for $R_{emulsion}$ is constructed as
\begin{equation}
\frac{R_{emulsion}}{w_1} = (0.270 + 0.0526 \frac{Q_m}{Q_i}) Ca_o^{-0.268}, 
\label{eq:EW}
\end{equation}            
where the parameter values 0.270, 0.0526 and -0.268 are obtained by fitting all the investigated periodic one-step data shown in figure \ref{fig:phasediagram} with the principle of minimum residual norm. To test the obtained scaling law, the values of the double emulsion radius $(R_{emulsion}/w_1)_{pred}$ computed from Eq. (\ref{eq:EW}) are plotted against the simulated radius values $(R_{emulsion}/w_1)_{simu}$ in figure \ref{fig:EW_IE} (a). The line of parity is plotted as a reference, and the closer the scattered data points are to the line of parity, the better the agreement is between the scaling law and the simulated results. It is seen that most of the points scatter around the line of parity, and the simple formula of Eq. (\ref{eq:EW}) can provide a general guidance for predicting double emulsion size.   

Another size of interest is the ratio between the equivalent inner droplet radius $R_{inner} = \sqrt{A_{inner}/\upi}$ and the entire double emulsion radius: $R_{inner}/R_{emulsion}$. \citet{Chang2009} experimentally proposed a scaling law for the double emulsion generated in co-axial capillaries. The inner droplet and the entire double emulsion are viewed to have the same formation time before being pinched off together in the dripping mode. According to the mass conservation law, $R_{inner}/R_{emulsion}$ is predicted by $R_{inner}/R_{emulsion} = (Q_i/(Q_i+Q_m))^n$, and the power-law exponent $n$ is 1/3 and 1/2 for three and two dimensional studies, respectively. Recently, \citet{Fu2016} numerically confirmed this relation in their two-dimensional study using a co-axial capillary device. However, the inner phase fluid actually breaks up slightly earlier than that of the middle phase fluid, especially in the current planar hierarchical flow-focusing device (see figure \ref{fig:evolution}). The difference in formation time between the two phases is also observed to be moderately affected by $Ca_m$ and $Ca_o$. To consider their effects, two power-law relations are assumed between $R_{inner}/R_{emulsion}$ and $Ca_m$ and $Ca_o$, respectively. A scale factor is also added to fine tune the entire size. 

Based on the above analysis, the scaling law for $R_{inner}/R_{emulsion}$ is constructed as
\begin{eqnarray}
\frac{R_{inner}}{R_{emulsion}} = 0.904(\frac{Q_i}{Q_i+Q_m})^{0.609} Ca_m^{-0.060} Ca_o^{0.030}. 
\label{eq:IE}
\end{eqnarray}  
The way to obtain the values of the coefficients in Eq. (\ref{eq:IE}) is the same to that used in Eq. (\ref{eq:EW}). The fitted power-law exponent of $Q_i/(Q_i+Q_m)$ is 0.609, which is close to 0.5 mentioned in the work of \citet{Chang2009} and \citet{Fu2016}. The difference can be attributed to the inconsistency in the breakup time of the inner and middle phases. Nevertheless, the difference in the formation time is small, which is also reflected by the scale factor 0.904 that is close to 1.0, and the near zero power-law exponents of $Ca_m^{-0.060}$ and $Ca_o^{0.030}$. Similar to figure \ref{fig:EW_IE} (a), the parity plot for the computed values of $(R_{inner}/R_{emulsion})_{pred}$ using Eq. (\ref{eq:IE}) and the measured values $(R_{inner}/R_{emulsion})_{simu}$ are shown in figure \ref{fig:EW_IE} (b). The good agreement between the scattered points and the parity line justifies the validity of the scaling law of Eq. (\ref{eq:IE}) for the $R_{inner}/R_{emulsion}$ values.

\subsection{Effect of interfacial tension ratio}

\begin{figure}
	\centering{\includegraphics[scale=0.41]{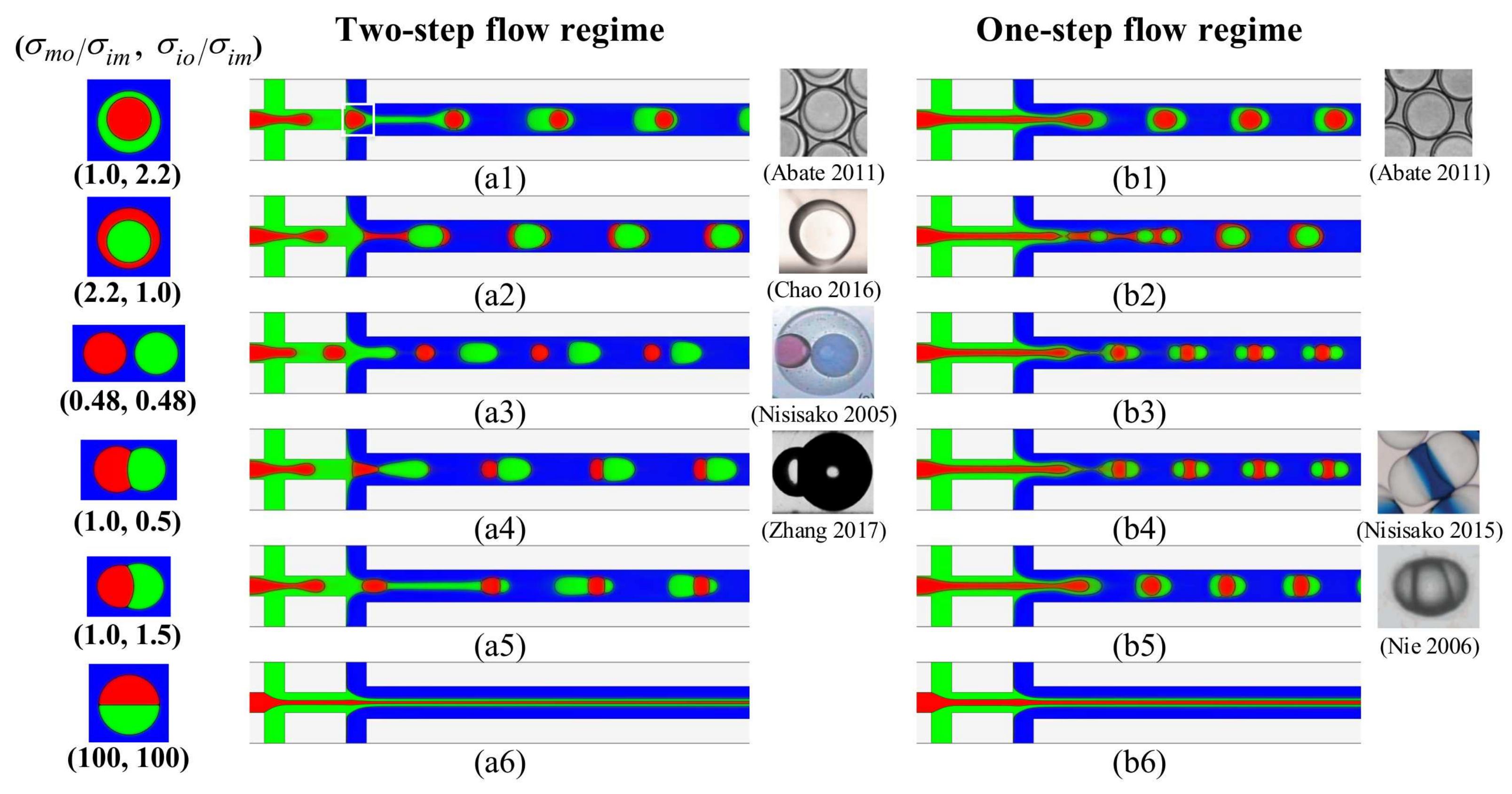}}
	\caption{Snapshots of emulsion formation behaviors under the effect of interfacial tension ratios in (a) two-step and (b) one-step formation regimes. The interfacial tension ratios for cases (a1, b1)-(a6, b6) are $(\sigma_{mo}/ \sigma_{im}, \sigma_{io} / \sigma_{im})$ = (1.0, 2.2), (2.2, 1.0), (0.48, 0.48), (1.0, 0.5), (1.0, 1.5) and (100, 100). The first column before (a) series shows the corresponding static equilibrium morphology of two equal-sized droplets at each interfacial tension ratio group. Relevant experimental works are put next to the related snapshots. The white square marked in (a1) highlights the region that the inner droplet is about to touch the middle-outer interface.}
	\label{fig:ITR}
\end{figure}

In figure \ref{fig:stability_diagram}, we show that a variation in the interfacial tension ratio could result in distinct equilibrium morphologies of two droplets of different fluids. To elucidate the role of interfacial tension ratios on the emulsion structure in different double emulsion formation processes, six groups of interfacial tension ratios that cover different regions of figure \ref{fig:stability_diagram} are investigated, i.e., $(\sigma_{mo}/ \sigma_{im}, \sigma_{io} / \sigma_{im})$ = (1.0, 2.2), (2.2, 1.0), (0.48, 0.48), (1.0, 0.5), (1.0, 1.5) and (100, 100) under two flow conditions for periodic two-step (Regime 1) and one-step (Regime 8) formation regimes. The flow parameters for the two-step and one-step formation regimes are given at $Ca_i$ = 0.012 and $Ca_i$ = 0.02, respectively, with $Ca_m$ = 0.011 and $Ca_o$ = 0.035. The corresponding flow rate ratios are $Q_i:Q_m:Q_o$ = 0.171 : 0.390 : 1 and 0.286 : 0.390 : 1. To obtain different interfacial tension ratios, $\sigma_{im}$ is fixed at 0.005 except for the case at $(\sigma_{mo}/ \sigma_{im}, \sigma_{io} / \sigma_{im})$ = (100, 100), where $\sigma_{im}=$ 0.00001 is used, similar to those used in figure \ref{fig:stability_diagram}. Figure \ref{fig:ITR} illustrates the snapshots of the (a) two-step and (b) one-step flow rates for each interfacial tension ratio group. Note that the first column before (a) series shows the corresponding static equilibrium morphology of each interfacial tension ratio group as shown in figure \ref{fig:stability_diagram}. Relevant experimental works are marked next to the related snapshots. 

It is seen in figure \ref{fig:ITR} that the formation details and the emulsion morphologies are greatly affected by the interfacial tension ratios in both formation regimes. Firstly, compared to the double emulsions obtained at $(\sigma_{mo}/ \sigma_{im}, \sigma_{io} / \sigma_{im})$ = (1.0, 2.2) (figure \ref{fig:ITR} (a1) and (b1)), the inverse engulfed double emulsion is captured in figure \ref{fig:ITR} (a2) and (b2) by reversing the interfacial tension ratios to $(\sigma_{mo}/ \sigma_{im}, \sigma_{io} / \sigma_{im})$ = (2.2, 1.0). With the inverse interfacial tension ratios, the inner phase fluid is more favored to the outer phase fluid and tends to engulf the middle phase droplet to lower the system's interfacial energy. In the two-step formation regime shown in figure \ref{fig:ITR} (a2), as the individually generated inner droplet approaches the second cross junction, it is getting closer to the middle-outer interface. Once the inner droplet touches the middle-outer phase interface, the attraction between the inner and outer phases would prompt the pinch-off of the middle phase layer between them and actuate the formation of the middle phase droplet. Afterwards, the inner droplet itself becomes a bridge connecting the newly formed middle phase droplet and the remaining middle phase front. Soon it breaks into two parts under the viscous force of the outer fluid. The inner phase portion adhered to the middle phase droplet evolves to wrap the middle phase droplet and the inverse double emulsion morphology is finally formed. In the one-step formation regime of figure \ref{fig:ITR} (b2), the inverse double emulsion is also obtained. However, the formation details are different due to the continuous supply of the inner phase fluid in the jetting mode. A string of small middle phase droplets are formed and connected by the inner phase fluid. The compound thread tip is then emulsified by the outer fluid for every two front middle phase droplets. The detached two middle phase droplets covered by the inner phase fluid soon merge with each other and produce a pure double emulsion.

\citet{Chao2016} experimentally captured the conversion from an initial double emulsion to its inverse structure using a glass-based capillary microfluidic device. Using the terminology of our work, an intermediate red-in-green-in-blue double emulsion is initially produced in their work, and the thermodynamic equilibrium green-in-red-in-blue configuration is only obtained after the external flow is stopped. However, in our work, the final configuration is formed directly without the intermediate red-in-green-in-blue configuration. This implies that the moment for interfacial tension dominating over the hydrodynamic effects in the formation behaviors is earlier in our simulations than that in the experimental work of \citet{Chao2016}. This could be explained by the experimental findings of \citet{Pannacci2008}. They pointed out that it is necessary for the inner droplet to touch the inner boundary of its host to evolve to thermodynamic equilibrium under the capillary forces. In other words, the sooner the three-phase contact line is formed, the faster the interfacial tension starts to dominate. For instance, if we look into the formation details in figure \ref{fig:ITR} (a2), there should be an instantaneous moment, like highlighted in the square region in figure \ref{fig:ITR} (a1), where the inner droplet is approaching the middle-outer interface due to the squeezing of the outer fluid. It allows the capillary force to act earlier. Regarding the experimental work of \citet{Chao2016}, a relatively thick middle layer surrounds the inner phase orifice in the co-axial glass capillaries, which could prevent the early formation of the three-phase contact line, and hence delay the interfacial tension effect.

At $(\sigma_{mo}/ \sigma_{im}, \sigma_{io} / \sigma_{im})$ = (0.48, 0.48), the red and green droplets tend to separate with each other at thermodynamic equilibrium. In the two-step formation regime shown in figure \ref{fig:ITR} (a3), the inner and middle phase droplets are successively formed and flow downstream without touching each other in the outer fluid, consistent with their static equilibrium morphologies. These alternately generated single droplets of two phase fluids have possible applications in being the source materials for producing multi-core emulsions \citep{Nisisako2005}. In figure \ref{fig:ITR} (b3), a more complex multiple emulsion is obtained in the one-step formation regime: an inner phase droplet is seized by two middle phase droplets on both sides in the flow direction. The contact length between the components of the multiple emulsion is decreasing when flowing downstream, but the components do not completely separate from each other in the finite computational domain. It can be attributed to two possible reasons. The first one is that the sequence structure results from a transient double emulsion rather than separately produced like in the two-step formation regime. Thus, it takes longer for the sequence structure to evolve to its thermodynamic equilibrium. Secondly, once the middle phase thread tip is pinched off, the lateral outer phase fluid rapidly fills the pinch-off region. Consequently, the most upstream component in the sequence is more accelerated and the hydrodynamic effects keep the three components staying next to each other. The complete separation of the components could be expected after the inflow pumps are stopped.

For the three cases shown in figure \ref{fig:ITR} (a4-a6) and (b4-b6), since the interfacial tensions in each case satisfy the Neumann triangle relation, the partial engulfing (Janus) emulsion should be achieved at thermodynamic equilibrium. \citet{Zhang2017From} experimentally captured the transformation from the core-shell structure to the Janus droplets based on prefabricated double emulsions. Here, our results in figure \ref{fig:ITR} (a4) show that the Janus droplet could be produced directly in the two-step formation regime within the same device for producing double emulsions. For figures \ref{fig:ITR} (a5), (b4), and (b5), biconcave and biconvex emulsions are formed downstream. These structures are analogous to those experimentally fabricated by \citet{Nisisako2015} and \citet{Nie2006}. Finally, for $(\sigma_{mo}/ \sigma_{im}, \sigma_{io} / \sigma_{im})$ = (100, 100) shown in figure \ref{fig:ITR} (a6) and (b6), $\sigma_{im}$ is so small that the inner and middle phase fluids can be approximately viewed as the same fluid, and the high $Ca_i$ induced by small $\sigma_{im}$ easily leads to the parallel layered flow behaviors for both the two-step and one-step flow conditions.     

\subsection{Effect of geometry}

Geometrical parameters in microfluidics are usually the key factors in single or double emulsion preparations \citep{Liu2011, Nabavi2015, Wu2017}. In this section, we focus on the effect of the geometrical parameters in changing the double emulsion formation regimes and the obtained double emulsion sizes. For the geometry shown in figure \ref{fig:geometry}, six normalized geometrical parameters can be defined as $w_2/w_1$, $w_3/w_1$, $w_4/w_1$, $w_5/w_1$, $w_6/w_1$ and $w_7/w_1$. Among them, the inlet length $w_6/w_1$ can be neglected, since the fully developed velocity distribution is always provided at the inlet, and the inner phase flow profile varies little before it reaches the middle phase inlet junction. Then, for simplicity, we make two assumptions to reduce the governing geometrical parameters, i.e., the side inlets for the middle and outer phase fluids have equal widths ($w_3 = w_2$), and the width of the channel connected the side inlets is set equal to that of the inner phase inlet ($w_4 = w_1$). Therefore, the main geometrical factors are reduced to the side inlet width ($w_2/w_1$), the main channel width ($w_5/w_1$) and the distance between the side inlets of the middle and outer phase fluids ($w_7/w_1$). Those geometrical factors are all investigated at two flow rates that lead to two-step and one-step formation regimes, respectively, for the original geometry. Different from the flow conditions used in the interfacial tension effect section, two closer $Ca_i$ values of 0.014 and 0.016 are used in this section at $Ca_m = 0.011$ and $Ca_o = 0.035$, to show the geometrical effect more obviously in changing the formation regimes. 
\begin{figure}
	\centering{\includegraphics[scale=0.37]{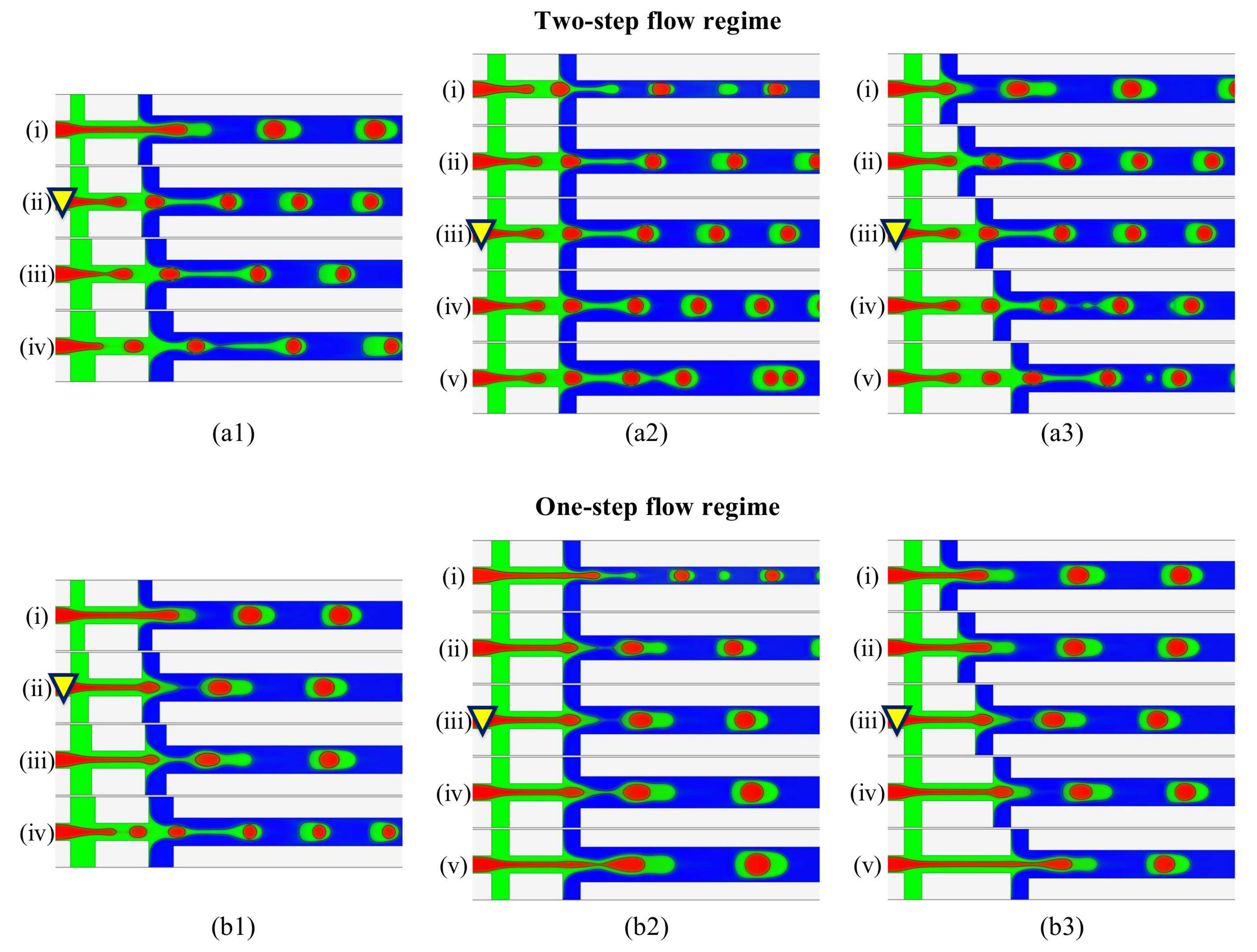}}
	\caption{Snapshots of double emulsion formation behaviors under the effect of geometrical parameters using the flow conditions that lead to (a) two-step and (b) one-step regimes in the original geometry. The results for the original geometry are marked with an inverted triangle. (a1, b1): $w_2/w_1$ ranges from 0.8, 1.0, 1.2 to 1.4 at $w_5/w_1$ = 1.6 and $w_7/w_1$ = 3.0; (a2, b2): $w_5/w_1$ ranges from 1.0, 1.4, 1.6, 1.8 to 2.0 at $w_2/w_1$ = 1.0 and $w_7/w_1$ = 3.0; (a3, b3): $w_7/w_1$ ranges from 1.0, 2.0, 3.0, 4.0 to 5.0 at $w_2/w_1$ = 1.0 and $w_5/w_1$ = 1.6.}
	\label{fig:effect of geometry}
\end{figure}

The effects of $w_2/w_1$, $w_5/w_1$ and $w_7/w_1$ on double emulsion formation behaviors are illustrated in figure \ref{fig:effect of geometry}. The (a) and (b) series correspond to the two-step and one-step flow rate conditions, and each parameter of concern increases from top to bottom in each sub-column. The results for the original geometry used in previous sections are marked with an inverted triangle. In figure \ref{fig:effect of geometry} (a1) and (b1), $w_2/w_1$ is increased from 0.8, 1.0, 1.2 to 1.4 at $w_5/w_1$ = 1.6 and $w_7/w_1$ = 3.0. It is seen that the breakup mode of the inner phase apparently changes from the jetting mode to the dripping mode with increasing $w_2/w_1$ at both flow rates. A larger $w_2/w_1$ is required to induce the inner breakup mode transition at higher $Ca_i$ values. Increasing the side inlet width increases the viscous force of the side-injected fluids to overcome the unaltered interfacial tension force, which leads to the breakup mode transition of the inner phase fluid. Figure \ref{fig:effect of geometry} (a1-ii)-(a1-iv) and figure \ref{fig:effect of geometry} (b1-i)-(b1-iii) illustrate the effect of increasing $w_2/w_1$ on emulsion sizes in the two-step and the one-step formation regimes. The size of the middle part increases in both formation regimes. However, the inner droplet size varies little in the two-step formation regime but decreases in the one-step formation regime.  

The effect of the main channel width is studied for $w_5/w_1$ = 1.0, 1.4, 1.6, 1.8 and 2.0 at $w_2/w_1$ = 1.0 and $w_7/w_1$ = 3.0 as shown in figure \ref{fig:effect of geometry} (a2) and (b2). It is seen that decreasing $w_5/w_1$ does not affect the formation regime of the inner phase fluid, but it could increase the breakup frequency of the middle phase fluid and induce the decussate regime, as observed in figure \ref{fig:effect of geometry} (a2-i, b2-i). For the flow-focusing geometry, all three inflow fluids converge to the main channel. Thus, narrowing the width of the main channel ($w_5$) increases the fluid velocity in the axial central region of channel, which creates a larger velocity gradient in the direction perpendicular to the main flow. During the expansion of the middle phase thread tip, it is subject to a higher shear stress, and as such the middle phase is more more likely to break up. With increasing $w_5/w_1$, the inner part and the entire double emulsion size vary little in the two-step formation regime (see figure \ref{fig:effect of geometry} (a2-ii)-(a2-iv)), but they both increase in the one-step formation regime (see figure \ref{fig:effect of geometry} (b2-ii)-(b2-v)). It indicates that the main channel width has a more obvious effect on the size of double emulsions generated in the one-step regime. Additionally, emulsions with two inner droplets are regularly obtained in the two-step formation regime at a wider collection channel, i.e., $w_5/w_1 = 2.0$ (see figure \ref{fig:effect of geometry} (a2-v)), similar to those experimentally captured in a double cross-junction device \citep{Deng2011} and capillary devices \citep{Nabavi2017M,Levenstein2016}.

At last, the distance between the two side inlets is investigated at $w_7/w_1$ = 1.0, 2.0, 3.0, 4.0 and 5.0, $w_2/w_1$ = 1.0 and $w_5/w_1$ = 1.6, as shown in figure \ref{fig:effect of geometry} (a3) and (b3). The two-step formation regime shifts to the one-step formation regime at $w_7/w_1$ = 1.0 (figure \ref{fig:effect of geometry} (a3-i)), where the inner phase front reaches the second junction before it breaks up in the dripping mode. However, more generally, the breakup modes and double emulsion sizes vary little with $w_7/w_1$ in both formation regimes, similar to the findings in binary systems using flow-focusing type geometries \citep{Wu2017, Utada2007}. Even though the lengthening of the connection channel increases the flow resistance through it, the flow behaviors inside vary little due to the slightly affected viscous force. As such, the velocities of the inner and middle phases are almost unaffected when they flow into the outer phase junction. Therefore, the overall flow behaviors are almost unchanged. Noteworthy, satellite droplets appear at $w_7/w_1 \geq 4.0$ in the two-step flow regime, due to the highly stretched middle phase thread tip during the emulsification process. It suggests that narrowing the distance between the side inlets could be a possible solution to avoid satellite droplets in producing double emulsions.

\section{Conclusions}

In this work, a two-dimensional ternary free energy lattice Boltzmann model is developed and used to systematically study the double emulsion formation behaviors in a planar hierarchical flow-focusing channel under variations of the flow rate, interfacial tension ratio and geometrical settings. 

The periodic two-step, one-step and decussate double emulsion formation regimes previously reported in the literature are qualitatively reproduced. A three-dimensional phase diagram is then constructed to show the distribution of each formation regime governed by $Ca_i$, $Ca_m$ and $Ca_o$ values. Depending on the breakup mode of the inner and middle phases, three distinct domains are classified as the periodic two-step, periodic one-step and non-periodic regions. The range for the periodic two-step region is almost unaffected by $Ca_o$, and it can be subdivided into seven formation regimes according to the pinch-off locations and the uniqueness of formation frequencies. Among them, periodic double emulsions are produced in Regime 1 and 2. In these two regimes, the entire double emulsion size decreases with $Ca_i$, increases with $Ca_m$, and varies little with $Ca_o$. Dripping-Threading regime (Regime 3) occurs when the middle phase fluid forms a continuous protective layer and carries multiple inner droplets. Decussate regimes (Regime 4) with one or two alternate empty droplets are both obtained. Noteworthy, the two empty droplets in the decussate regime could be produced either in a one-by-one sequence, or by breaking an initially formed large empty droplet into two daughter droplets. The bidisperse behaviors in double emulsion size and formation frequency are captured in a certain range of $Ca_i$ values in the two-step formation regime. The bidispersity could exist simultaneously for both the inner and middle phase fluids (Regime5), or only occur to the inner phase fluid (Regime 6 and 7). In the periodic one-step region for double emulsions (Regime 8), the entire double emulsion size is found to decrease with $Ca_i$ and $Ca_o$, but increases with $Ca_m$. Compared to the two-step formation regime, $Ca_o$ has a more obvious effect on the size of double emulsions formed in the one-step regime. Based on the one-step data (Regime 8), two empirical scaling laws are constructed for the size of the entire double emulsion and the proportion of the inner droplet. The good predictions of both scaling laws justify that the one-step formation process of double emulsions can be analogously viewed as a sum of a blocking period and a squeezing period. 

Another contribution of this work is that the presented free energy model is capable of dealing with a wide range of interfacial tension ratios, and provides accurate results for predicting complete engulfing double emulsions, partial engulfing Janus droplets and non-engulfing separate droplets. In particular, it was necessary to include an additional free energy term to capture the complete engulfing double emulsions. In the current microfluidic device, a variation in the interfacial tension ratios leads to distinct emulsion morphologies, including the inverse engulfing double emulsions \citep{Chao2016}, non-engulfing single droplets \citep{Nisisako2005}, Janus droplets \citep{Zhang2017From}, biconcave and biconvex emulsions \citep{Nisisako2015, Nie2006}, and even parallel flows.

Regarding channel geometrical parameters, the breakup mode of the inner phase fluid is changed from dripping to jetting by decreasing the side inlet width $w_2/w_1$, or by narrowing the distance between the two phase side inlets $w_7/w_1$. This leads to the conversion from the two-step formation regime to the one-step formation regime. The main channel width $w_5/w_1$ should not be too small in order to avoid the decussate regime. Moreover, narrowing $w_7/w_1$ is a possible solution to get rid of the satellite droplets for double emulsions generated in the two-step regime. The entire double emulsion size increases with $w_2/w_1$, but is rarely affected by $w_5/w_1$ or $w_7/w_1$ in the two-step formation regime. For the one-step formation regime, the double emulsion size increases with $w_2/w_1$ and $w_5/w_1$, but is independent of $w_7/w_1$.   

We would like to point out that the above work is carried out in a two-dimensional scheme. The present ternary free energy model could be directly extended to three dimensions. The main differences lie in the spatial and velocity discretization schemes, which we have resolved e.g. in \citet{Sadullah2018} and \citet{Wohrwag2018}. Based on the fundamental knowledge achieved in the present work, a three-dimensional study is the next step. Indeed, we mostly obtain the jetting regime of the inner phase fluid at $We_i \sim 1.0$ with the current 2D ternary LBM, in contrast to the dripping regime obtained by the reference studies. We believe the main reason for this difference could be attributed to the 3D effects. Since the dispersed thread grows faster at larger $We_i$, a stronger capillary effect is needed to promote droplet formation in the dripping regime \citep{Utada2005,Fu2012}. It means that the Laplace pressure contribution induced by the out-of-plane curvature plays an important role to promote droplet breakup at larger $We_i$ \citep{Hoang2013}, which is absent in 2D simulations. In addition, wall confinement in the out-of-plane direction can also become important. For example, \cite{Azarmanesh2016} pointed out that the double emulsion formation behaviors in flow-focusing channels are also related to the pressure buildup at the upstream of the inner phase fluid, which is influenced by both the in-plane and out-of-plane wall confinements.

It will be interesting to generalize the scaling laws presented here to three dimensions, and to compare them against experimental observations. We expect the forms of the scaling laws in Eqs.\eqref{eq:EW} and \eqref{eq:IE} to remain the same but different exponents will be obtained in three dimensions. Furthermore, equal density fluids are used at present. Our newly developed high-density ternary free energy model \citep{Wohrwag2018} could be applied to investigate double emulsion formation behaviors with other fluid types, where density difference is an important factor. It is also worth extending the current ternary free energy model to deal with multiple emulsions with more components (N $>$ 3), or introducing variable interfacial tensions governed by the surfactants \citep{Liu2018} to study more complex fluid systems.    

\section*{Acknowledgments}
H. Liu, C. Zhang and N. Wang acknowledge financial supports from the National Key Research and Development Project of China (No. 2016YFB0200902), and the National Natural Science Foundation of China (Nos. 51876170, 51506168 and 51711530130). H. Kusumaatmaja acknowledges funding from EPSRC (No. EP/P007139/1). C. Semprebon acknowledges support from Northumbria University through the Vice-Chancellor's Fellowship Programme, and funding from EPSRC (No. EP/S036857/1). N. Wang was supported by the China Scholarship Council for one year (2017-2018) at Durham University, UK. 

\section*{Declaration of Interests}
The authors report no conflict of interest.

\appendix
\section{}\label{appA}

The expressions for the additional terms in the chemical potentials due to the additional energy term in Eq. \eqref{eq:Ea} are provided below:
\begin{equation}
\left\{
\begin{array}{llll}
C_1<0: &\mu_\rho= \beta\left(\frac{\rho+\phi-\psi}{2}\right), 
&\mu_\phi= \beta\left(\frac{\rho+\phi-\psi}{2}\right), 
&\mu_\psi=-\beta\left(\frac{\rho+\phi-\psi}{2}\right); \\[2pt]
C_2<0: &\mu_\rho= \beta\left(\frac{\rho-\phi-\psi}{2} \right),  
&\mu_\phi=-\beta\left(\frac{\rho-\phi-\psi}{2} \right), 
&\mu_\psi=-\beta\left(\frac{\rho-\phi-\psi}{2} \right); \\[2pt]
C_3<0: &\mu_\rho= 0, 
&\mu_\phi= 0, 
&\mu_\psi=2\beta\psi; \\[2pt]
\end{array} \right.
\end{equation}
\begin{equation}
\left\{
\begin{array}{llll}
C_1>1: &\mu_\rho= \beta\left(\frac{\rho+\phi-\psi}{2}-1 \right), 
&\mu_\phi= \beta\left(\frac{\rho+\phi-\psi}{2}-1 \right), 
&\mu_\psi=-\beta\left(\frac{\rho+\phi-\psi}{2}-1 \right); \\[2pt]
C_2>1: &\mu_\rho= \beta\left(\frac{\rho-\phi-\psi}{2}-1 \right),  
&\mu_\phi=-\beta\left(\frac{\rho-\phi-\psi}{2}-1 \right), 
&\mu_\psi=-\beta\left(\frac{\rho-\phi-\psi}{2}-1 \right); \\[2pt]
C_3>1: &\mu_\rho= 0, 
&\mu_\phi= 0, 
&\mu_\psi=2\beta(\psi - 1); \\[2pt]
\end{array} \right.
\end{equation}

\bibliographystyle{jfm}
\bibliography{jfm_double_emulsion}

\end{document}